\documentstyle[12pt,psfig]{article}
\newcommand{\aj}[1]{{\sl AJ}, {\bf #1}}
\newcommand{\apj}[1]{{\sl ApJ}, {\bf #1}}

\newcommand{\apjsupp}[1]{{\sl ApJS}, {\bf #1}}
\newcommand{\aan}[1]{{\sl A\&A}, {\bf #1}}

\newcommand{\pasp}[1]{{\sl PASP}, {\bf #1}}

\newcommand{\mnras}[1]{{\sl MNRAS}, {\bf #1}}

\begin {document}
\begin{center}
\Large\bf{$^{56}$Ni dredge-up in the type~IIp Supernova~1995V}
\end{center}
\vspace*{0.5cm}
\large{A.Fassia$^{1}$, W.P.S. Meikle$^1$, T.R. Geballe $^2$, 
N.A. Walton$^3$,D.L. Pollacco$^3$,\\
\sf\large R.G.M. Rutten$^3$, C. Tinney$^4$ \\
\\ 
\small
$^1$Astrophysics Group, Blackett Laboratory, Imperial College, Prince Consort Rd
, London SW7 2BZ, UK\\
$^2$Joint Astronomy Centre, 660 N. A'Ohoku Place, University Park,
Hilo, Hawaii 96720, USA \\
$^3$Royal Greenwich Observatory, Apartado de Correos 321, 38780 Santa
Cruz de La Palma, Tenerife, Islas Canarias, Spain \\
$^4$Anglo-Australian Observatory, PO Box 296 Epping, NSW 2121, Australia \\
\begin{abstract}
We present contemporary infrared and optical spectra of the plateau
type~II SN~1995V in NGC~1087 covering four epochs, approximately 22 to 84 days
after shock breakout. The data show, for the first time, the {\it
infrared} spectroscopic evolution during the plateau phase of a typical
type~II event. In the optical region P~Cygni lines of the
Balmer series and of metals such as Sc~II, Fe~II, Sr~II, Ca~II and
Ba~II lines were identified. The infrared (IR) spectra were largely
dominated by the continuum, but P~Cygni Paschen lines and Brackett~$\gamma$
lines were also clearly seen. The other prominent IR features are confined
to wavelengths blueward of 11000~\AA\ and include Sr~II 10327, Fe~II
10547, C~I 10695 and He~I 10830~\AA.  Helium has never before been
unambiguously identified in a type~IIp supernova spectrum during the
plateau phase.  We demonstrate the presence of He~I 10830~\AA\ on days
69 and 85.  The presence of this line at such late times implies
re-ionisation.  A likely re-ionising mechanism is $\gamma$-ray
deposition following the radioactive decay of $^{56}$Ni. We examine
this mechanism by constructing a spectral model for the He~I 10830~\AA\
line based on explosion model s15s7b2f of Weaver \& Woosley (1993).  We find 
that this does not generate the observed line owing to the confinement
of the $^{56}$Ni to the central zones of the ejecta. In order to
reproduce the He~I line, it was necessary to introduce additional
upward mixing or ``dredge-up'' of the $^{56}$Ni, with $\sim$10$^{-5}$ of
the total nickel mass reaching above the helium photosphere.  In
addition, we argue that the He~I line-formation region is likely to
have been in the form of pure helium clumps in the hydrogen envelope. 
The study of He~I 10830~\AA\ emission during the photospheric phase of
core-collapse supernovae provides a promising tool for the constraint
of initial mixing conditions in explosion models.
\end{abstract}
\hspace*{0.3cm}
\vspace*{0.2cm}

\newpage
\section{Introduction}
Type~IIp (plateau) supernovae form the classic subgroup of the
core-collapse supernovae.  They are believed to arise from massive
stars (12-25 M$_{\odot}$) during the red supergiant phase. Early
theoretical work by Falk and Arnett (1973) showed that hydrodynamical
instabilities should appear in explosions of such massive stars.  As
the shock wave propagates through the stellar envelope it sets up
density and pressure profiles which can in some cases result in the
formation of Rayleigh-Taylor~(RT) instabilities (Chevalier 1976).  In
particular, RT instabilities are expected to grow at the interface of
the core and the hydrogen envelope because of the large entropy (and
density) jump that occurs there (Weaver \& Woosley 1980).  A direct
consequence of these instabilities is that chemical mixing in the
ejecta takes place (Bandiera 1984).

Herant \& Woosley (1994) have studied 2-D simulations of red supergiant
explosions over a wide mass range, and found that the growth of
hydrodynamic instabilities is highly likely in all cases.  They showed
that as the explosion (outgoing) shock plows into the hydrogen
envelope, a reverse (ingoing) shock is formed, and between them, RT
instabilities grow. Bubbles of hydrogen formed by these instabilities
are violently dragged towards the centre of the star by the reverse
shock. Simultaneously, compact helium and oxygen clumps advance out
into the hydrogen envelope and bubbles of $^{56}$Ni are formed and
distributed in the outer parts of the ejecta. It has also been realised
that strong dredge-up should result from the neutrino-driven
convection close to the neutron star surface (Herant \& Benz 1992)
which has been invoked to account for the conversion of the
core-collapse to explosion. This should also produce fast-moving blobs
or ``fingers'' of radioactive material which eventually penetrate the
outer layers of the supernova.

So far, only SN~1987A has provided us with clear observational evidence
for dredge-up in a core-collapse event. This includes the shape of the
light curve, the early detection of X-rays and $\gamma$-rays and the
width of the iron lines in the infrared.  However SN~1987A was, of
course, only a single event, and a rather unusual one in that it arose
from a blue supergiant progenitor. Therefore, we cannot simply assume
that similar dredge-up occurs in all other type~II events. Indeed,
simulations have shown that differences in progenitor structure can
lead to significantly modified hydrodynamical evolution (Herant \& Benz
1992; Herant \& Woosley 1994).  Clearly, to establish whether or
not deep dredge-up is typical of all core-collapse supernovae, a major
step would be to demonstrate dredge-up and mixing in the most-common of
all core-collapse events, the type~IIp supernova.

A powerful demonstration of the occurrence of deep dredge-up would be
the appearance of radioactive material at the surface at early times.
Helium lines arising in the supernova envelope can be used as a tracer
of the upwardly-mixed radioactive material. Helium lines are of high
excitation. During the first week of the supernova, recombination
maintains the populations of the excited levels and so He~I lines are
seen.  However after about 10 days, the conditions in the type~IIp
atmosphere are such that all the helium will have recombined and
de-excited to the ground state.  But, if dredge-up occurs during the
explosion, radioactive $^{56}$Ni may reach the outer parts of the
supernova envelope at early times.  If it does, the $\gamma$-rays from
its decay ($^{56}$Ni $\Rightarrow$ $^{56}$Co $\Rightarrow$ $^{56}$Fe)
will excite or re-ionise the helium. Thus the detection of helium lines
during the plateau phase (20-120~d post explosion) should imply upward
mixing of radioactive material from the core.  Unfortunately no optical
He~I lines have ever been unambiguously identified during the plateau
phase. However, there are two well-known strong lines in the infrared
viz. He~I 10830~\AA\ (2s$^{3}$S--2p$^{3}$P$^{0}$) and 20580~\AA\
(2s$^{1}$S--2p$^{1}$P$^{0}$) which offer the prospect of testing for
dredge-up of radioactive material. This technique was applied by Graham
(1988) and Chugai (1991) using the 10830~\AA\ line in the SN 1987A at
early times. Lucy (1991) invoked upward mixing of $^{56}$Ni to account
for strong optical He I lines in the early-time spectra of type Ib
supernovae.

To investigate dredge-up in type~IIp supernovae we began in 1995 a
programme of infrared and optical spectroscopy of this type of
supernova. The data we present here comprise an extensive set of
IR/optical spectroscopic observations of the type~IIp SN~1995V,
spanning epochs of 22 to 85 days post explosion.  The observations are
described in section 2.  In section 3 we compare the observations with
a simple spectral synthesis model and discuss the line identifications,
especially in the IR. In section 4 we describe the method we used to
estimate the amount of dredge-up of $^{56}$Ni. In section 5 we present
the results from the comparison of the model with the data, and our
estimations for the amount of dredge-up. In section 6 we discuss the
implications of this work for our understanding of dredge-up.

\section{Observations}

SN~1995V was discovered in NGC~1087 on 1995~August~1 by R.~Evans
(1995). It lay 21$^{\prime\prime}$ east and 3$^{\prime\prime}$  south
of the centre of its host galaxy. The unfiltered CCD magnitude of the
supernova at discovery was +15 (Dopita \& Trung Hua 1995) and strong
P~Cygni Balmer lines observed on August 2 (Benetti 1995) showed it to
be a type~II event near maximum light. Below, we deduce that to within
a couple of days, the epoch of shock breakout was 1995~July~25.  We adopt
this as epoch 0~days and refer all other epochs to this fiducial date.
Subsequent to its discovery, no broad-band photometry for SN~1995V was
reported.  However, we can be certain that it was a {\it plateau}
type~II event since the spectra exhibit a broad, well-developed
absorption component at H$\alpha$ (Figure~1).  This is generally absent
from {\it linear} type~IIs (Schlegel 1996).  Identification of SN~1995V 
with a plateau type~II event is confirmed by the evolution of the 
optical flux, as explained below.

\subsection{Optical observations} 

The optical spectroscopy observations of SN~1995V were carried out
using the Intermediate Dispersion Spectrograph (IDS) of the Isaac Newton
Telescope (INT) on La Palma.  Spectra were obtained on days 23, 34, 72
and 84 post-shock breakout (see below). The optical spectroscopy log is
given in Table~1.  

The spectra were reduced by means of standard routines in the
data-reduction package FIGARO (Shortridge 1991). The CCD frames were
debiased, flat-fielded, and sky-subtracted.  The spectra were extracted
using the optimal-extraction algorithm of Horne (1986). Wavelength
calibration was by means of CuNe and CuAr arc lamps.  Feige~25 was the
spectrophotometric flux standard. The optical spectra are presented in
Figure~1.  The fluxes shown in Figure~1 have been scaled as explained
below.

Observing conditions were poor during most of the optical observations,
and only on day~23 were conditions close to photometric.  On this day,
we judge the fluxing to be accurate to $\pm$15\%.  The fluxing for the
other three epochs is less certain.  Nevertheless, in order to confirm
that SN~1995V was a type~IIp event, we derived ``BV magnitudes'' from the
optical spectra by multiplying them with the filter response functions
B3 (also known as $\phi$B) and V of A\v{z}usienis \& Strai\v{z}ys
(1969). These magnitudes are presented in Table~2.  Note that these
magnitudes are uncorrected i.e. they were derived {\it before}
application of the flux scaling corrections described below.

We then compared the B~magnitudes with those of the type~IIp SN~1983K 
(Phillips {\it et al.} 1990) (Figure~2).
Unusually, this supernova was discovered before maximum light 
and so its time of shock break-out, t$_0$, is known
quite accurately. In Figure~2 it can be seen that, in spite of the
fluxing uncertainties for SN~1995V, the B~magnitudes show a similar
evolution to those of SN~1983K during its plateau phase. 

The horizontal
displacement between the two light curves is constrained by the fact
that SN~1995V was not detectable with a 0.41m reflector on 25 July
1995 (Evans, 1995). The similarity of the light curve
shapes confirms that SN~1995V was indeed a plateau type~II. From the
comparison of the two light curves, we deduce that the time of the
shock break-out, t$_0$, for SN~1995V was 25 July 1995 $\pm$2 days.

\subsection{Infrared observations} 
Infrared spectra were obtained at the Anglo-Australian Telescope (AAT),
Siding Spring, using the Infrared Imaging Spectrograph, IRIS, and at
the United Kingdom Infrared Telescope (UKIRT), Hawaii, using the Cooled
Grating Spectrograph, CGS4.  Spectra were obtained on days 22, 42, 44, 69
and 85.  The IR spectroscopy log is given in Table~3.  

Given the
complexities of instrument scheduling on major telescopes, the match in
epoch between the four optical and four IR spectra is remarkable. 
Note that these data are the first ever infrared spectra obtained of a
type~IIp during its photospheric phase. The IRIS spectra were reduced
using FIGARO.  The cross-dispersed spectral orders were first
straightened. The frames were then flat-fielded and sky-line
subtracted.  The spectra were then optimally extracted.  Wavelength
calibration was performed by means of a CuAr arc, and the flux standard
was HD~19904, assuming J=6.727. The CGS4 data were reduced using CGS4DR
(Daly \& Beard 1992) and FIGARO.  A krypton arc was used for wavelength
calibration, and BS~770 and HD~18881 were the flux standards.  The
adopted magnitudes were J=5.60, H=5.4, K=5.35 for BS770 and J=7.13,
K=7.14 for HD18881. The IR spectra are shown in Figure~3. The fluxes
shown in Figure~3 have been scaled as explained below.

\subsection{Flux calibration} 
Photometric conditions occurred for the IR observations on days~22 and 69, 
and so we judge the IR fluxing at these
epochs to be better than $\pm$10\%.  However, for the other epochs the
fluxing of the optical and IR spectra presented problems owing to both adverse
weather conditions and technical difficulties during the observations.
In order to improve the fluxing accuracy for those spectra taken under 
difficult conditions, we proceeded as follows.  The
day~72 optical spectrum was scaled so that the flux at the red end was
consistent with that of the blue end of the well-fluxed day~69 IR spectrum,
assuming a Rayleigh-Jeans slope in the small (360~\AA) gap between the
optical and IR coverage.  The scaled optical spectrum was then used to
derive a corrected V~magnitude using the method described above. A
template V-band light curve was derived from the observations of the
type~IIp SNe 1969L (Ciatti, Rosino \& Bertola 1971) and 1991G (Blanton
{\it et al.} 1995). The SN~1995V day~72 corrected V~magnitude was
then used to set the template zero-point for this supernova, and the
resulting light curve used to obtain the V~magnitudes for days 23, 34
and 84.  The optical spectra were then scaled to match these
V~magnitudes.  For day~23 this spontaneously provided an excellent
match to the well-fluxed day~22 IR spectrum in the overlap region, 
reinforcing our
belief that the scaling procedure was valid. The lack of IR light curves
for this type of supernova meant that a different approach was required
to correct the IR fluxing on days 42, 44 and 85. Correction of the
JH-band spectrum on day~44 was achieved using a linear interpolation of
the IR fluxes between days~22 and 69, while the day~42 K-band spectrum
was scaled to match the overlap with the corrected day~44 JH-spectrum. 
The day~85 IR spectrum was simply scaled to match the day~84 optical
spectrum in the overlap region.

\section{Line identifications}

The optical spectra shown in Figure~1 are consistent with SN~1995V
being a classic type~IIp event.  The H$\alpha$ P~Cygni profile is
strong throughout the plateau phase.  Other prominent lines include
H$\beta$, Na~I-D (possibly blended with He~I~5876~\AA) and the Ca~II~IR
triplet. As SN~1995V evolved the absorption line profiles, especially
those of H$\alpha$, H$\beta$ and the Ca~II triplet, became narrower and
deeper. The number of discernible lines increased with time so that by
day~72 it was very difficult to determine the location of the continuum
level in the B and V bands.  As expected (Branch 1987; 
Elias {\it et al.} 1988), the IR~spectra (Figure~3)
have fewer features than in the optical region, but these
also became more pronounced as the supernova evolved.  Especially
prominent is a group of lines in the 10000--11200~\AA\ range.  In
addition, Paschen lines and Br~$\gamma$ are all present.  The
identification of other IR lines is discussed below.

To investigate the spectra we used a simple spectral model where we
assumed a spherical, homologously expanding envelope. We adopted a
power law density profile, $\rho(r)=\rho(ph)(\frac{r}{r_{ph}})^{-n}$,
where $\rho_(ph)$, $r_{ph}$ are the density at, and the radius of the
photosphere respectively.  The model continuum consisted of a dilute
blackbody which incorporated the flux dilution ($\zeta$) parameters
specified by Eastman {\it et al.} (1996).  The temperature was
adjusted to match the observed continuum.  The lines were formed in a
pure-scattering atmosphere lying above the dilute blackbody
photosphere (Castor 1970).  For each transition, we assumed that the
population of the upper level is negligible compared with that of the
lower level.  (Support for this assumption was provided by our
detailed modelling of the He~I~10830~\AA\ line described below, where
the lower level population was found to exceed that of the upper level
by a factor of about 10$^4$). Line blending was treated using the
formulae for doublet profiles described by Castor {\it et al.} (1979).
For the line identifications, line IDs of Williams (1987), Elias {\it
et al.}(1988), McGregor {\it et al.} (1988) and Meikle {\it et al.}
(1989) and the predicted line strengths of Branch (1987) were taken
into account.  Transition wavelengths and oscillator strengths were
taken from Kurutz (1996). The models were reddened using A$_V$
calculated from the empirical wavelength dependence formula of
Cardelli {\it et al.}  (1989). The absolute visual extinction,
A$_V$=1.37$\pm$0.11, was estimated from the interstellar Na I-D lines,
which were observable in all the optical spectra. This was done using
the relation E(B-V)$\sim$ 0.25 EW (Na~I-D) which Barbon {\it et al.}
(1990) derived from their supernova observations, together with an
assumed A$_V$=3$\times$E(B-V) (Whitford 1958). The model was also
redshifted to match the recession velocity of NGC~1087,
v$_0$=1519$\pm$6~km/s (de Vaucouleurs {\it et al.} 1991).

For each line, the model's free parameters were the population at the 
photosphere of the
lower level of the transition, N$_{low}(ph)$, and the
velocity at the photosphere, v(ph. An additional free parameter
was the density profile index, $n$, where the lower level population at radius r is given by: $N_{low}(r)~=~N_{low}(ph) (\frac{r}{r_{ph}})^{-n}$. 
At a given epoch, the weak metal lines, such as those of Fe~II, Ba~II
and Sc~II all tended to yield the same values for v(ph) and $n$.
In Table~4 we show these values.  Their uncertainties are
judged to be less than $\pm$10\%.  The constancy of v(ph) amongst
these lines at a particular epoch, together with their weakness,
indicate that they were optically thin and that the derived velocities
correspond to material at the true photosphere.  For the
optically-thin lines, we find that v(ph) declined from 6200~km/s on
day~22 to 2200~km/s on day 85.  Compared with SN~1987A, SN~1995V
presented higher photospheric velocities at comparable epochs.  The
density index obtained from the weak lines was then adopted for the
model matches to the optically-thick lines.  Significantly higher
velocities were found for the prominent, strong lines of hydrogen and
the calcium IR triplet and so we conclude that they were optically
thick throughout the 22--85~day period.

The modelled and observed spectra are compared in Figures~4 (optical)
and 5 (infrared). Line identifications, together with their individual
values for N$_{low}(ph)$ and v(ph), are presented in Tables 5 and
6. For optically thick lines, ``v(ph)'' corresponds to the velocity
at which the line became optically thick. 

In the optical region, lines of neutral hydrogen, carbon, sodium and
calcium, together with lines of singly-ionised carbon, calcium,
scandium, iron, strontium and barium were identified. At later epochs,
the matches to the data in the blue region are less satisfactory due to
line blanketing.  Consequently, the line identifications in this
wavelength region on days 72 and 84 are less secure. The model also
fails to reproduce the emission component of H$\alpha$.  This effect
has been known about for many years.  It is explained by invoking
collisional excitation from the n=2 to n=3 level which greatly enhances
the emission component (e.g. Branch {\it et al.} 1981). A similar
effect is not seen in the other Balmer lines since the line optical
depths are such that, rather than escape, the line photons are more
likely to be converted to H$\alpha$ plus a Paschen, Brackett etc line (
Branch {\it et al.} 1981; Xu {\it et al.} 1992).  This also adds to the
H$\alpha$ emission.

The infrared spectra are largely dominated by the continuum. As
indicated above, Paschen lines and Brackett $\gamma$ are also present
at all epochs. The other infrared lines include Sr~II 10327~\AA, Fe
II 10547~\AA, C I 10695~\AA\ and He~I 10830~\AA.  As
can be seen from Figure 5 the model underproduces the emission
components of the Paschen lines i.e. the IR hydrogen lines are clearly
not formed by pure resonant scattering. Similar behaviour was observed
in the early-time spectra of SN~1987A (e.g. Larson {\it et al.} 1987,
Meikle {\it et al.} 1989) and is thought to be due to the conversion
of Balmer line photons to H$\alpha$ plus Paschen, Brackett, etc line
photons, as mentioned above.

Of special interest is the identification 
of the Sr~II~10327~\AA\ line. 
The line was also identified in the SN 1987A spectra (Elias {\it et al.} 1988 )
The presence of this line and of the Ba II lines indicates that s-process 
element enhancements might have occurred in the supernova progenitor
(Williams 1987).

To double-check the He~I~10830~\AA\ identification we repeated the
model fit but with the helium line excluded.  The result is shown in
Figure 6.  Clearly the no-helium model provides a much inferior match
to the data.  We conclude that He~I~10830~\AA\ was definitely present
on days~69 and 85. {\it We believe that this is the first time that
helium has been clearly identified in a type~IIp supernova during the
plateau phase.} The velocity at the He~I~10830~\AA\ ``photosphere''
declined from 4250~km/s to 3750~km/s between days~69 and 85.  This is
substantially higher than the corresponding photospheric velocities of
the weak lines and even those of the Balmer lines (apart from
H$\alpha$), indicating that the helium was optically thick.  The
absence of He~I~20580~\AA\ does not cast doubt upon the
He~I~10830~\AA\ identification since our model indicates that the 
He~I~20580~\AA\ line is at least a factor of 100 weaker.

\section {Investigation of $^{56}$Ni dredge-up using the He~I
10830~\AA\ line.}

The He~I 10830~\AA\ (2s$^{3}$S-2p$^{3}$P$^{0}$) line is formed in a
region where the population of the metastable 2s$^{3}S$ level is
maintained by the balance of the recombination rate of He$^+$ with the
rates of collisional de-excitation and forbidden radiative decay 
(Elias {\it et al.} 1988). However, because the He$^+$ recombination
time scale is $<$ 1 d (Graham 1988) we expect that at the times
observed (69d-85d) all helium should be neutral and in its ground
state. The persistence of the He~I line at such late times requires an
increasing ionisation rate to raise the electron density faster than it
is diluted by the expansion.  Following
Graham (1988), to account for the persistence of the He~I line, we
propose that the ionisation is maintained by the decay of $^{56}$Co
which resulted from the decay of $^{56}$Ni produced in the explosion.
The decay $\gamma$-rays and resultant Compton-scattered electrons
re-ionize the helium.  The subsequent recombination produces the
observed He~I 10830~\AA\ line.

To explore the radioactive decay scenario, we calculate the radioactive
energy deposited in the outer envelope of the supernova and hence find
the ionisation balance as a function of radius. We then determine the
2s$^{3}S$ population density profile.  This is then used to replace the
2s$^{3}S$ power-law profile in the P~Cygni models described above
and the calculated
spectrum compared with the observations.  We first applied this technique 
using a specific explosion model, s15s7b2f (Weaver \& Woosley 1993).  We 
considered two cases {\it viz.} one in which the helium in the envelope is
entirely microscopically mixed with the hydrogen (i.e. 100\% primordial
H/He), and one in which various fractions of the helium in the envelope
are clumped.  As will be explained below, the presence of clumps of
unmixed helium above the photosphere can dramatically reduce the
required degree of $^{56}$Ni dredge-up. However, it will be shown that
even for the clumped-helium case, model s15s7b2f does not reproduce the
He~I 10830~\AA\ line.  We then modified model s15s7b2f by increasing
the extent of the dredge-up of $^{56}$Ni to produce a match to the
observations.  Again, we consider cases where helium is unclumped and
clumped.

The decay of $^{56}$Ni and $^{56}$Co produces a number of $\sim$1~MeV
$\gamma$-ray lines of different energies. The energy is deposited in
the expanding envelope via Compton-scattering. The differential
scattering cross-section and the fractional energy transfer are highly
dependent upon the scattering angle. Moreover, the high-energy
electrons produced by the Compton scattering go through many
interactions with ions and electrons before finally yielding up the
last of their energy (Sutherland \& Wheeler 1984). Ideally, such a
process should be modelled using a Monte Carlo simulation in order to
compute the rate at which the high energy electrons transfer their
energy to the environment.  However, Monte Carlo techniques are
excessively demanding of computational resources.  For this initial
study, therefore, instead of attempting a Monte Carlo analysis, we
adopted the ``gray'' approximation of Sutherland and Wheeler (1984). In
their method, they assume a purely absorptive $\gamma$-ray opacity,
$\kappa_{\gamma}$, which is independent of the energy of the
$\gamma$-rays. The value they adopt, $\kappa_{\gamma}$=0.03~cm$^{2}$g$^{-1}$, is in accord
with that based upon Monte Carlo simulations (Colgate $\it {et ~al.}$
1980, Sutherland \& Wheeler 1984, Swartz $\it {et ~al.}$ 1995).  The
details of the deposition calculation are included in the Appendix.

In a H/He envelope of low ionisation, all the energy produced by the
Compton scattering of the $\gamma$-rays is deposited in the H and He
causing ionisation, rather than in the heating of the 
electron gas (Meyerott 1980).
Therefore, for a deposition rate of $\epsilon_{i}$ in
species i the abundance of the next ionisation stage i+1 is given by
the energy balance : \begin {equation}
n_{i+1}=\frac{\epsilon_{i}}{n_{e} \alpha_{i} w_{i}} \end {equation}
where $n_{e}$ is the electron density ($n_{e}=n_{He~II}+n_{H~I}$),
$\alpha_{i}$ is the recombination rate to all levels of the species i 
and $w_{i}$ is the energy required to produced an ion-electron pair. The ratio
of energy deposited in H and He is $\epsilon_{He}/\epsilon_{H}=mY$,
where $Y=n_{He}/n_{H}$.  From the Bethe-Bloch formula for energy loss of
fast electrons in H/He material, we find $m$ to be 1.7.

Our model includes the detailed effect of the fast electrons on the
populations of excited states of He~I. We ignore direct excitation
since the fast electrons excite only singlet states, which decay
rapidly back to the ground state. The key populating process is
recombination.  Of the recombinations to excited levels of He~I,
approximately three-quarters are to triplet states, with the remainder
going to singlet states (Osterbrock 1989). Recombinations to singlet
levels ultimately populate the 2s$^{1}$S and 2p$^{1}$P levels.  Atoms
in the 2p$^{1}$P state decay mostly to 1s$^{1}$S but some may also
decay to 2s$^{1}$S. However the 1s$^{1}$S-2p$^{1}$P transition is
highly optically thick with the result that all the singlet
recombinations eventually pass through the 2s$^{1}$S state.  Atoms in
2s$^{1}$S level decay by two-photon emission (A=51~s$^{-1}$) to the
ground state. Thus, only a small population of excited singlet states
is built up and consequently the optical depths of absorption lines
arising from excited singlet levels are small.

All captures to triplet levels lead, through downward radiative
transitions, to the highly metastable 2s$^{3}$S level where the
population can become quite substantial. There are a number of
mechanisms which can depopulate this level. Two, purely radiative,
depopulating mechanisms can be identified. A very weak single-photon
radiative decay to 1s$^{1}$S can occur (A=1.27$\times10^{-4}$
s$^{-1}$) (Osterbrock 1989).  Of considerably greater significance,
however, is the 2-stage intersystem radiative decay
2s$^{3}$S$\rightarrow$2p$^{3}$P$\rightarrow$1s$^{1}$S. This occurs in
the presence of a radiation field (such as from the photosphere),
which is required to excite the first stage. There are also two
important depopulating processes involving collisions. One of these
simply involves thermal electron collisions causing excitation or
de-excitation from 2s$^{3}$S across to singlet states.  The other
process is Penning ionisation (Bell 1970, Chugai 1991):

\begin {equation}
He~I~ (2^{3}S)+ H \rightarrow He~I~ (1^{1}S) + H^{+} + e.
\end {equation}

This process requires microscopic mixing of hydrogen and helium but in
the event of such mixing, it dominates the depopulation of the
2s$^{3}$S level (see below).

We can therefore summarise the population balance of the 2s$^{3}$S 
level with the equation:

\begin{equation}
n_{He(2^{3}S)} (n_{e}Q + C_{P} + R + A)= \alpha(n^{3}L) n_{e}n_{He~II}.
\end {equation}

$Q$ is the sum of the collision rates from 2s$^{3}$S to all singlet
states, and has the value 1.826$\times$10$^{-8}$cm$^{3}$s$^{-1}$ (Berrington
\& Kingston  1987). $C_{P}$ is the Penning ionisation rate and is given
by \begin{math} C_{P} =\gamma_{P} n_{H} \end{math} where
\begin{math}\gamma_{P}= 7.5\times 10^{-10}(\frac{T}{300K})^{1/2}
\end{math} cm$^{3}$s$^{-1}$ (Bell 1970, Kozma 1996). For the
recombination temperature of hydrogen (T$\sim$5000 K), which we adopted
for the days concerned (plateau phase),
$\gamma_{P}$=3$\times$10$^{-9}$cm$^{3}$s$^{-1}$. $A$ =
1.27$\times10^{-4}$ s$^{-1}$, is the A-value 
for the single-photon radiative decay to 1s$^{1}$S.  $\alpha(n^{3}L)$ =
3.26$\times$ 10$^{-13}$cm$^{3}$s$^{-1}$ is the total recombination
coefficient for the triplet states at a temperature of 5000~K
(Osterbrock 1989). R is the two stage
inter-system radiative decay rate (Chugai 1991). This is given by:

\begin{equation}
R=B_{23}\frac{4\pi}{c}(\frac{D}{vt})^{2} F_{\nu}^{C}e^{0.92 A(\lambda)}
A_{32}^{-1} A_{31} \beta_{31}
\end{equation}
where $\beta_{31}$ is the escape probability for the photon emitted in
the decay 2p$^{3}$P$\rightarrow$ 1s$^{1}$S and $F_{\nu}^{C}$ is the flux
of the continuum at $\lambda$=10830~\AA, determined from the observed
infrared spectra. A($\lambda$) is the extinction calculated using the
empirical formula by Cardelli {\it et al.} 1989 ({\it cf.} Section 3). For the
distance D, a value of 16.7 Mpc was derived from the infall model described
in Aaronson {\it et al.} 1982. $v$ is the velocity at the He~I
``photosphere'' at time t ({\it cf.} Table 6)

As mentioned above, the population balance is extremely sensitive to
the effects of Penning ionisation. This is illustrated in Table 7 where
we compare the 2s$^{3}$S depopulation rates in microscopically mixed
H/He for the different mechanisms. The rates were calculated for
$n_{e}\sim10^{6}cm^{-3}$, $n_{H}\sim3\times10^{10}cm^{-3}$ and
$n_{He}\sim4\times10^{9}cm^{-3}$, which are typical for the He~I
10830~\AA\ line-formation region at the epochs concerned (t$\sim$
60-80 days).  We see that Penning ionisation dominates the depopulation
of the 2s$^{3}$S level. Thus, for typical conditions, any helium that is
intimately mixed with hydrogen will have its 2s$^{3}$S state so heavily
depopulated that its contribution to the 2s$^{3}$S population in the
envelope will be negligible.  Putting it another way, pure helium
clumps above the 2s$^{3}$S photosphere, such as might be produced by
RT instabilities at the H/He interface, will tend to
dominate the 2s$^{3}$S population.  Helium that is microscopically
mixed with hydrogen, such as primordial helium, will make a negligible
contribution even if its abundance is much greater.  Thus, the
$^{56}$Ni-driven ionisation rate inferred from the He~I 10830~\AA\
data depends on what fraction of the helium in the envelope is clumped
and what fraction is intimately mixed with the hydrogen.  The greater
the microscopic mixing, the greater must be the $^{56}$Ni-driven
ionisation rate to produce the observed He~I 10830~\AA.  Although it
is likely that RT instabilities at the H/He interface will
mix some clumps of helium up into the hydrogen layer (Herant \& Woosley
1994), unfortunately the actual fraction of clumped helium in the
envelope is unknown. We therefore examine two cases - one involving no
clumps of pure helium, and one involving two different degrees of
helium clumping.

\subsection {He~I 10830~\AA\  line strength predictions from s15s7b2f 
explosion model}
\subsubsection{Case 1: Hydrogen and helium 100\% microscopically mixed}

In this case the He~I 10830~\AA\ line is formed entirely by
microscopically mixed (primordial) helium, and so the dominant
mechanism for the depopulation of the 2s${3}$S level is Penning
ionisation. Equations (1) therefore reduce to :

\begin{equation}
\frac{\epsilon_{H}}{w_{H}}+n_{He(2^{3}S)}n_{H~I}C_{P}=n_{H
II}n_{e}\alpha_{H~I}
\end{equation}
\begin{equation}
\frac{\epsilon_{He}}{w_{He}}=n_{He~II}n_{e}\alpha_{He}
\end{equation}

where \begin{math}\epsilon=\epsilon_{He}+\epsilon_{H}\end{math} is the
energy deposited in the envelope.  From the Bethe-Bloch formula
$\epsilon_{He}/\epsilon_{H}=mY$, and assuming $n_{e}=n_{He II} +n_{ H
II}$ we can calculate the population of the 2s$^{3}$S level,
$n_{2s^{3}S}$(R) as a function of radius, R, using
equations (5) and (6). This requires the energy deposition rate
$\epsilon(R)$, the total number density
$n_{tot}(R)=n_{He~I}(R)+n_{H~I}(R)+n_{e}(R)$, and the relative
abundance of hydrogen and helium, $Y=\frac{n_{He II}(R)+n_{He
I}(R)}{n_{H II}(R)+n_{H I}(R)}$. The last two quantities were obtained from explosion
model s15s7b2f. 

In this model, a 15 M$_{\odot}$ progenitor is exploded, ejecting 0.0643
M$_{\odot}$ of $^{56}$Ni. The density profile of the $^{56}$Ni goes to
zero at 750~km/s from the core. Using this density profile we calculated
the deposition rate $\epsilon(R)$, using the formula of Sutherland \& Wheeler
(1984) as described in the Appendix.  The calculated density profile of
the population of the 2s$^{3}$S level, $n_{2s^{3}S}$(R), was then fed
into the P~Cygni models for days 69 and 85.  These are presented in
Figure~7 compared with the observations. Clearly the contribution
of He~I 10830~\AA\ to the model spectrum is negligible ({\it cf.}
Figure~6). We deduce that the $^{56}$Ni density profile specified by
model s15s7b2f is inadequate to account for the observed He~I 2s$^{3}$S
population if 100\% microscopic mixing of the helium and hydrogen is
assumed.

\subsubsection{Case 2: Clumps of helium in the hydrogen envelope}
If clumps of ``pure'' helium exist in the hydrogen envelope then, as
explained above, the observed He~I 10830~\AA\ line will be formed
almost entirely by the clumped helium since depopulation of 2s$^{3}$S
by Penning ionisation will not affect such regions. We investigated the
effect of helium clumping by assuming that a certain fraction of the
helium mass in a region was in the form of clumps {\it i.e.} unmixed
with hydrogen.  The exclusion of the Penning mechanism in this case
meant the main de-populating mechanisms were 2-stage intersystem
radiative decay and collisional depopulation.  In order to calculate
the effects of these processes on the 2s$^{3}$S population it was
necessary to use a more detailed helium model consisting of the ground
state 1s$^{1}$S and the first four excited states. The
processes considered in our model included recombination, collisional
excitations and de-excitations, radiative decays (allowing for the escape probability of the line photons), and the two-stage
radiative transition 2s$^{3}$S$\rightarrow$2p$^{3}$P$\rightarrow$1s$^{1}$S
for the depopulation of the 2s$^{3}$S level.  Energy levels and
empirical formulae for collision strengths were provided by Berrington
\& Kingston (1987) for T=5000~K. Radiative decay rates were provided by
Mendoza (1983) and recombination coefficients for each level were
obtained from Osterbrock (1989).

The electron density in the clumps was estimated from the balance
between ionisations and recombinations:
\begin {equation}
\frac{\epsilon^{clumps}}{\omega_{He}}=n_{He~II}^{clumps}
n_{e}^{clumps} \alpha_{He}
\end{equation}
where $\epsilon^{clumps}$ is the energy deposited per sec per cm$^{3}$
in the clumps, $\omega_{He}$ is the energy required to produce an ion
pair, and $\alpha_{He}$ is the recombination coefficient for He$^+$.
Because the clumps consist of pure helium, it follows that
$n_{e}^{clumps}=n_{He^{+}}^{clumps}$. We can therefore estimate the
electron density in the clumps using equation (7), provided we know the
energy deposited there.

As explained in the Appendix, the energy per sec per cm$^{3}$ that is
deposited in a region is proportional to the density of the region {\it
i.e.} $\epsilon \propto \rho$.  We envisage that the pure helium clumps
are in pressure balance with a surrounding envelope which has the
primordial H/He abundance ratio .  Consequently we have:
\begin{displaymath}
P_{clumps}=P_{env.}  
\Longrightarrow
\frac{ \rho_{clumps}N_{Av}kT}{\mu_{He}}=
\frac{\rho_{env.}N_{Av}kT}{\mu_{env}} \Longrightarrow
\end{displaymath}
\begin{equation}
\rho_{clumps}=\rho_{env.}\frac{\mu_{He}}{\mu_{env}} 
\end{equation}
where $N_{Av}$ is Avogadro's number, k is Boltzmann's constant, and
$\mu_{env}$, $\mu_{He}$ are the mean molecular weight of the envelope
gas and the helium clumps respectively. The mean molecular weight of the
mostly neutral gas in the He~I line-forming region is given by:
\begin{math} \mu_{env}\approx(\frac{X_{H}}{1}+
\frac{X_{He}^{'}}{4}+\frac{X_{Z}}{2})^{-1}\end{math} where $X_{H}$,
$X_{He}^{'}$, $X_{Z}$ are the abundances of the hydrogen, primordial
helium, and metals in the envelope gas. We assumed that a certain
fraction, $\chi_{He}$, of the total helium mass is clumped i.e. unmixed
with hydrogen. Therefore the residual primordial helium abundance,
$X_{He}^{'}$, becomes: \begin{math}X_{He}^{'}= X_{He}-(\chi_{He}
X_{He})\end{math}, where $X_{He}$ is the average helium abundance in
the zone considered. The radial profiles of the abundances $X_{H}$,
$X_{He}$, $X_{Z}$ were obtained from the s15s7b2f explosion model.
In practice, $X_{Z}$ was small and so was omitted from the calculation.

Using equation (8) with $\chi_{He}$=0.5, we
calculated the corresponding clump densities and then calculated the
energy deposited there. We then determined the 2s$^{3}$S level
population in the clumps using the five-level helium atom model
described above. The calculated population profile, $n_{2s^{3}S}$(R),
was fed into the P~Cygni model and the resulting model spectra
compared with the observations. These are presented in Figure~7.
In spite of the fact that the clumped helium is unaffected by Penning
ionisation, we still find that the He~I 10830~\AA\ makes a negligible
contribution to the model spectrum, even for $\chi_{He}$ as high as
0.5. We conclude that even by invoking clumped helium, explosion model
s15s7b2f is still unable to yield sufficiently high 2s$^{3}$S level
populations to account for the observed helium line. Given that
$^{56}$Ni only extends to 750~km/s in model s15s7b2f, this failure is
not surprising since, for example, at 69~days the helium photosphere is
at 4250~km/s. The gamma-rays simply do not penetrate to sufficiently
high velocities.  {\it We conclude that the radial extent of $^{56}$Ni
must be greater than specified by model s15s7b2f}.

\subsection{ He~I 10830~\AA\ line strength predictions from model
s15s7b2f plus additional $^{56}$Ni dredge-up}

We propose that the persistence of the He~I 10830~\AA\ line is due to
the upward mixing of $^{56}$Co into the outer layers of the supernova
envelope.  A similar proposal was made by Graham (1988) for the
peculiar type~II SN~1987A.  To investigate the extent of this upward
mixing, we have modified the s15s7b2f model results so as to place
radioactive nickel beyond 750~km/s. 

To do this, we invoked a $^{56}$Ni density profile in which the density
remained constant up to velocity v$_{c}$ and then declined as a power
law of index $k$ (Figure 8). The total $^{56}$Ni mass was always held
at the s15s7b2f value {\it viz.} 0.0643 M$_{\odot}$.  v$_{c}$ and $k$
are free parameters. The density profiles for other elements in
s15s7b2f were unchanged.  In each zone the nickel was homogeneously
mixed with the other elements. We then proceeded as before to calculate
the 2s$^{3}$S population profile.  This was then applied to the P~Cygni
model and the model spectrum compared with the observations. $k$ and
v$_{c}$ were adjusted to provide simultaneous matches to the observed
spectra on days 69 and 85.  As before, we considered cases where the
helium was unclumped or clumped.  The ranges of values for v$_{c}$ and
$k$ which provided good matches are shown in Table 8, and examples of
corresponding $^{56}$Ni profiles are shown in Figure 9. An example of
the synthetic spectra produced using the values $\chi_{He}$=0.1, k=9,
v$_{c}$=800km/s is shown in Figure 10. 

Clearly, with sufficient upward mixing of the $^{56}$Ni, the modified
s15s7b2f model is able to reproduce the He~I 10830~\AA\ emission.  We
note that in the case of clumped helium, v$_{c}$ has roughly the same
value as the $^{56}$Ni boundary in model s15s7b2f viz. $\sim$750~km/s.
However, for the case where the helium is entirely microscopically
mixed with the hydrogen we find, unsurprisingly, that v$_{c}$ must
extend much further viz. to $\sim$1150~km/s. In the power law zone, we
find an index of about 8 or 9 for all the cases considered. For each
case, we calculated the $^{56}$Ni mass in the power law zone, and the
mass above the helium photosphere. These are presented in Table 8.  In
all cases, we see that about a third of the total $^{56}$Ni mass must
have been dredged up beyond the v$_{c}$ limit.  The larger v$_{c}$ in
the no-clumped helium case implies a considerably greater dredge-up.
Dramatically different masses of $^{56}$Ni above the He~I photosphere
are inferred for the clumped and no-clumped cases, with a factor of
$\sim$30 more $^{56}$Ni being required for pure microscopic mixing,
relative to 10\% clumping.

\section{ Discussion}
Reproduction by explosion model s15s7b2f of the observed He~I emission
is achieved only by invoking substantial additional dredge-up of the
$^{56}$Ni from the core.  As expected, the case with no helium clumping
requires the greatest dredge-up. However, we argue that the no-clumping
case is probably highly unrealistic, since it is difficult to see how
one could achieve such a large amount of $^{56}$Ni dredge-up and yet
have no pure helium clumps in the same environment.  We deduce,
therefore, that there must exist some pure helium clumps in the
hydrogen envelope.  As we have shown, with fairly modest clumping of
$\chi_{He}=$0.1 or 0.2, the uniform central core coincides, to within
the errors with that of the unmodified s15s7b2f model.  The addition of
a steep, power-law density component to this will bring sufficient
$^{56}$Ni to the surface to account for the He~I 10830~\AA\ emission.
Nevertheless, the fraction of the $^{56}$Ni mass which must be dredged
up beyond the uniform core is quite substantial. For SN~1995V we
conclude that a) a small amount of $^{56}$Ni ($\sim$10$^{-6}$ M$_{\odot}$)
must have been dredged up to the helium photosphere
(v$\sim$4,250~km/s), and b) clumps of pure helium must have also
existed in this region.

High velocities for the decay products of $^{56}$Ni ($\sim$ 3000 km/s)
were also observed in the ejecta of SN 1987A (e.g. Meikle et al. 1993).
As shown by Herant \& Benz (1992), if the $^{56}$Ni is located at the
base of the ejecta at t$\sim$300s it is impossible to accelerate even a
small fraction to about 3000 km/s during subsequent instabilities.  In
order to achieve such high velocities, it is necessary to invoke
outward mixing of the nickel at even earlier times, such as might be
caused by neutrino convection. If this occurred, then the later
instabilities would carry the nickel to still higher velocities. In
order to match the observations of SN1987A, Herant \& Benz had to
premix nickel out to 1.5 M$_{\odot}$ above the mass cut. 

Herant \& Woosley (1994) studied shock propagation,
mixing and clumping in the explosion of red supergiants.  In order to
take into account the pre-mixing of $^{56}$Ni during the initiation of
the explosion, they diluted the nickel by a factor $\sim$4 above the
mass cut. They then followed the shock propagation and the growth of
RT instabilities.  For all progenitors their simulations
showed that extensive RT instabilities develop in the
ejecta in the wake of the reverse shock from the H/He interface. In
contrast to the blue supergiant studies, these instabilities have ample
time in which to evolve and completely reshape the ejecta. In spite of
this, in all the explosions simulated, nickel did not reach velocities
higher than $\sim$1500 km/s. Similarly, helium did not exceed
velocities higher than 2500 km/s. Our results, therefore, indicate that
a higher degree of pre-mixing may be required than is invoked in the
Herant \& Woosley models.

Recently, Bazan \& Arnett (1997) have simulated mixing in core-collapse
events. Their simulations include both RT and
Richtmeyer-Meshkov instabilities.  These produce much higher
velocities for $^{56}$Ni than do R-T instabilities alone. Velocities as
high as $\sim$4000~km/s are predicted. The mass and profile index of
the upwardly-mixed $^{56}$Ni derived above could be of considerable
value in constraining the initial parameters of these instabilities.

\section{Conclusions}

He~I 10830~\AA\ has been clearly identified in the infrared spectra
of SN~1995V during the plateau phase. The velocity of the helium
``photosphere'' derived from this line declined from 4250 km/s to
3750km/s between days 69 and 85. This was substantially higher than the
corresponding velocities of the metal lines and even of the Balmer
lines indicating that helium was optically thick.  The presence of the
He~I line at such late times suggests re-ionisation driven by the decay
of $^{56}$Co.

Using explosion model s15s7b2f we were unable to reproduce the observed
He~I line.  However, with the additional upward mixing (dredge-up) of
$\sim$10$^{-6}$ M$_{\odot}$ of $^{56}$Ni to high velocities we were able
to match the observed spectra. We also deduce that the He~I
line-forming region took the form of clumps of pure helium in the
hydrogen envelope.  We note that these deductions are insensitive to the 
accuracy of the flux calibration since they are based primarily on the optical depth
of the He~I P~Cygni line.  We conclude that the He~I 10830~\AA\ line can be
used as a valuable probe for the study of mixing in a type~II
supernova.  The next stage will be to extend this study to provide
constraints on the size of the initial perturbations in mixing models.

\vspace{\baselineskip}
{\large\bf Acknowledgements:}  We thank Brian Schmidt for helpful
discussions, and in particular for originally pointing out to us the
need for infrared spectroscopy of type~IIp supernovae in the plateau
phase.  We also thank Stan Woosley for his advice and for the use of
his explosion model. We are also grateful to Ron Eastman, Marcos Montes
and Phil Pinto for useful discussions. We thank Charlene Heisler and 
Mike Irwin for carrying 
out some of the observations.  AF is supported by a scholarship
from the Alexander S Onassis Public Benefit Foundation.  Much of the
data reported here were obtained via the AAO, ING and UKIRT Service 
Programmes.
\\

\newpage
\bf\Large{References}\\
\\
\normalsize
Aaronson M., Huchra J., Mould J., Schechter P.L., Tully R.B., 1982,
\apj{258} ,64 \\
A\v{z}usienis A., Strai\v{z}ys V., 1969, Soviet Astr., {\bf 13}, 316\\
Ambwani K., Sutherland P., 1988, \apj{325}, 820 \\
Bandiera R., 1984, \aan{139}, 368 \\
Barbon R., Benetti S., Rosino L., Cappellaro E., Turatto M., 1990,
\aan{237}, 79B \\
Bazan G., Arnett, D., 1998 \apj{496}, 316 \\
Bell K.L., 1970, J.Phys.B:Atom.Molec.Phys. {\bf 3}, 1308 \\
Benetti S.,1995, IAU Circ.6197 \\
Berrington K. A., Kingston A. E.\, 1987, J. Phys. B: At. Mol. Phys.,
{\bf 20}, 6631 \\
Blanton E.L., Schmidt B.P., Kirshner R.P., Ford C.H., Chromey
F.R.,~Herbst W., 1995, \aj{110}, 2868 \\
Branch B., Falk S.W., McCall M.L., Rybski P., Uomoto A.K, Wills B.J.,
1981, \apj{244}, 780 \\
Branch D., 1987, \apj{320}, L121 \\
Cardelli J., Clayton G., Maths J., 1989, \apj{345}, 245 \\
Castor J.I., 1970, \mnras{149}, 111\\
Castor J.I., 1979, \apjsupp{49}, 481 \\
Chevalier R.A., 1976, \apj{207}, 872 \\
Ciatti F., Rosino L., Bertola F., 1971, Mem.Soc.Astrom.Ital,{\bf 42},
163 \\
Chugai N.N., 1991, in Supernovae, ed. Woosley S.E., 
Springer-Verlay, New York, p286 \\
Colgate S.A., Petschek A.G., Kriese J.T., 1980, \apj{237}, L81 \\
Daly P.N., Beard S.M., 1992, Starlink User Note 27.1 \\
De Vaucouleurs G., De Vaucouleurs A., Corwin JR. H. G., Buta R. J., Paturel G.,
Fouquet E., 1991, Third Refernce Catalogue of Bright Galaxies, Version
3.9 \\
Dopita M., Trung H.C., 1995, IAU Circ.6197 \\
Eastman R.G., Schmidt B.P., Kirshner R.P., 1996, \apj{466}, 911\\ 
Elias J.H., Gregory B., Philips M.M., Williams R.E., 
Graham J.R., Meikle W.P.S., Schwartz R.D., Wilking B., 1988, \apj{331},
L9 \\ 
Evans R.0., 1995, IAU Circ.6197 \\
Falk S.W., Arnett D.W., 1973, \apj{180}, L65 \\
Graham J.R., 1988, \apj{335}, L53 \\
Herant M., Woosley S.E., 1994, \apj{425} , 814 \\
Herant M., Benz W., 1992, \apj{387}, 294 \\
Horne K., 1986, \pasp{98}, 609 \\
Kozma C., 1996, PhD Thesis, Department of Astronomy, Stockholm
University \\
Kurutz R.L., 1996, Atomic spectral line database, \\
http://cfa-www.harvard.edu/amp/data/kur23/sekur.html \\ 
Larson P.H., Drapatz S., Mumma M.J., ESO Conference and Workshop
Proceedings No26 1987, 147 \\
Lucy L.B., 1991, \apj{383}, 308 \\
Meikle W.P.S., Allen D.A., Spyromilio J., Varani G.-F., 1989,
\mnras{238}, 193 \\
Meikle  W.P.S., Spyromilio J., Allen D.A., Varani G.-F., Cumming R.J.,
1993, \mnras{261}, 535 \\
Mendoza C., 1983, in Planetary Nebulae, ed. Flower D.R., p143 \\
Meyerot R.E., 1980, in Supernovae Spectra, ed. R.E. Meyerott and
G.H. Gillespie (AIP Conf. Proc. 63), p49 \\
McGregor P., 1988 Elizabeth and Frederick White Conference on Supernova 1987A, 
Canberra 1988, June 20-24, Proc.astr. Soc. Aust. 7,450 \\
Osterbrock D.E., 1989, in Astrophysics of Gaseous Nebulae and Active
Galactic Nuclei, p25 \\
Philips M.M., Hamuy M., Maza J., Ruiz T., Carney B.W., Graham J.A.,
1990, \pasp{102}, 299 \\
Schlegel E.M., 1996, \aj{111}, 1660 \\
Shortidge K., FIGARO General data reduction and Analysis Styarlink MUD,
RAL, June 1991 \\
Sutherland P.G., Wheeler J.C., 1984, \apj{280}, 282 \\
Swartz D.A., Sutherland P.G., Harkness R.P., 1995, \apj{446}, 766 \\
Weaver T.A., Woosley S.E., 1980, in AIP Conf. Proc. 63 ,Supernovae
spectra, ed R.Meyerott, G.H. Gillespie (New York: AIP), 15 \\
Weaver T.A., Woosley S.E., 1993, Phys. Rep.{\bf227}, 65 \\
Whitford A.E., 1958, \aj{63}, 201 \\ 
Williams R.E., 1987, \apj{320}, L117 \\ 
Xu Y., McCray R., Oliva E., Randich S., 1992, \apj{386}, 181 \\
\newpage
\appendix
\Large\bf{Appendix: Calculation of the deposition energy}\\
\\
\normalsize
Our calculation of the $\gamma$-ray deposition function is based on the formula
described by Sutherland \& Wheeler (1984). The formula is basically an
absorption calculation which assumes that the $\gamma$-ray opacity,
$\kappa_{\gamma}$, is independent of energy and totally absorptive.
The value adopted, $\kappa_{\gamma}=$0.03cm$^{2}$g$^{-1}$, is in accord
with that based upon Monte Carlo simulations (Colgate {\it et al.}
1980, Sutherland \& Wheeler 1984, Swartz {\it et al.} 1995).

The geometrical aspects of the absorption calculation were treated as
follows. Consider an arbitrary point A in the ejecta, of mass $dm=\rho dAdl$,
where $dA$ is an infinitesimal area normal to the direction to a source
of $\gamma$-rays at some other point. Then the rate of the energy deposited at
this point due to the source is:

\begin {equation}
ds_{dep}=\kappa_{\gamma} \rho dl \frac{dA}{4 \pi R^{2}}e^{-\tau(\bf{r})}
s(\bf{r})
\end {equation}
where we take $dm$ to be the origin of the co-ordinates, and the source
at $\bf{r}$, emitting $\gamma$-ray energy at the rate $s(\bf{r})$
ergs g$_{rad}^{-1}$s$^{-1}$. The source $s(\bf{r})$ is spatially
constant but it varies with time due to the exponential decay
of the daughter nuclei. The total rate of the deposition function is
given by the dimensionless deposition function $d\alpha$ obtained by
dividing $ds_{dep}$ by $dm$ and  $s(\bf{r})$ = constant and by integrating over the
radioactive source region. Therefore we have:

\begin {equation}
d\alpha=\frac{\kappa_{\gamma}}{4\pi}\int d^{3}r \frac{1}{r^{2}} e^{-\tau(\bf{r})}
\rho_{rad}(\bf{r}) 
\end {equation}
where $\rho_{rad}(\bf{r})$ is the density of the radioactive material
and $\tau(\bf{r})$ is the optical depth,
\begin {equation}
\tau(\bf{r})=\int_{0}^{\bf r }\kappa_{\gamma} \rho(s) ds
\end {equation}
$\rho$ is the total density. Assuming spherical symmetry
equation (10) becomes:

\begin{eqnarray*}
da=\frac{\kappa_{\gamma}}{2} \int d\mu \int dr 
exp[-\tau(r)]
\rho_{rad}(r) \Rightarrow \\
da(r_{A})=\frac{1}{2}\int d\mu \int_{A}^{C} exp[-\tau] f_{rad}(r) \rho
\kappa_{\gamma} dr =
\end {eqnarray*}
\begin {equation}
\frac{1}{2}\int d\mu \int_{A}^{C} d\tau f_{rad}(r)exp[-\tau]
\end {equation}

\begin{figure}
\centerline{\psfig{figure=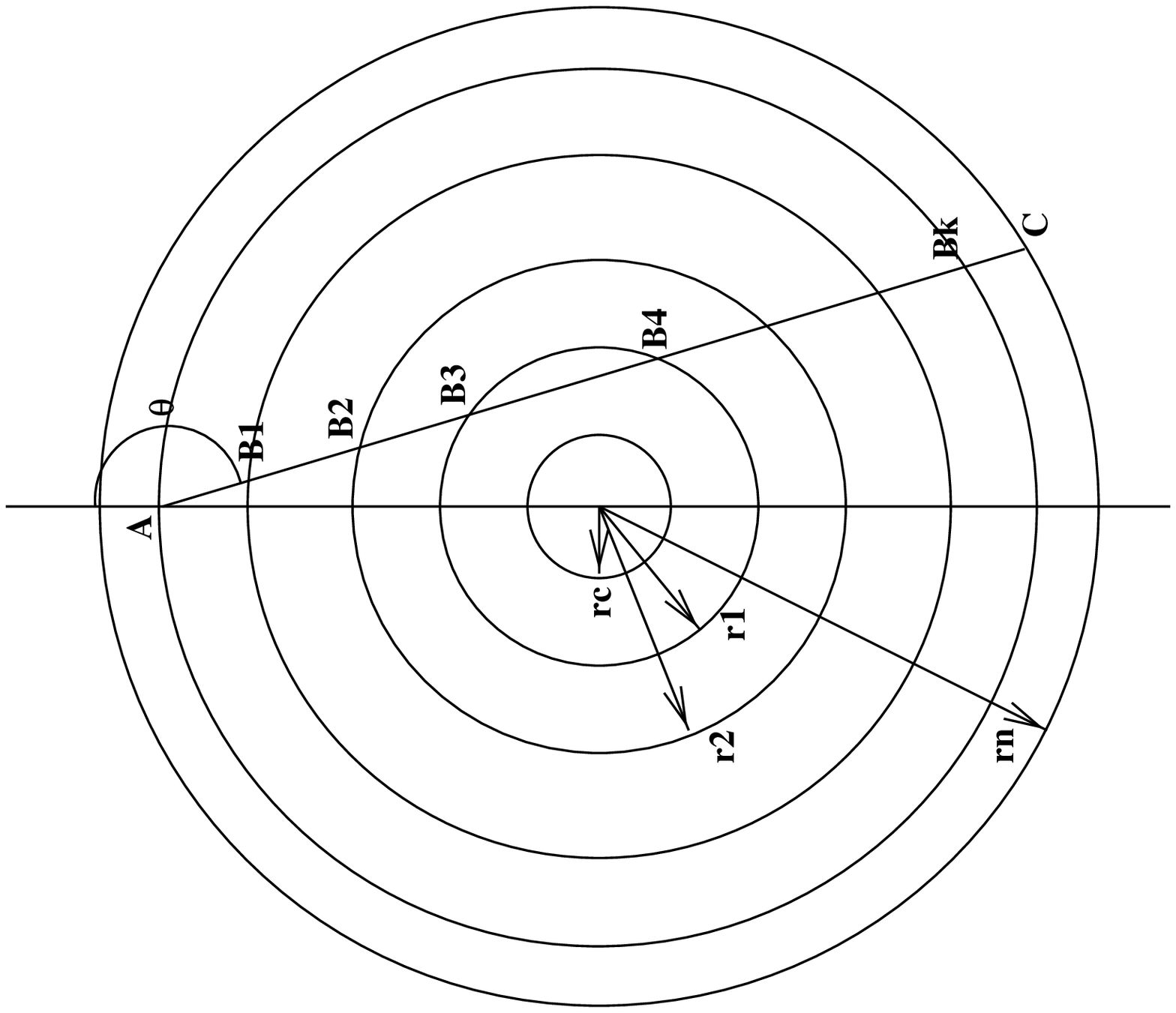,width=2.7in,height=3.in,angle=270}}
\vspace*{0.3cm}
\hspace*{0.5cm} {\bf Figure A:} Geometry for the calculation of the energy
deposition by $\gamma$ rays 
\end{figure}

where $f_{rad}(r)$ is the fraction of the $^{56}$Ni mass relative to the total
 mass and $\mu=cos\theta$ ({\it cf.} Figure A). For the calculation of the
 dimensionless deposition 
function $da$ we used the density profile of the s15s7b2f explosion
 model (Weaver \& Woosley 1993).  The model is divided in shells 
within each of which the
density and the initial mass fraction of nickel, $f_{rad}$ are
 constant. Equation (12) can then be integrated exactly ({\it cf.}
 Figure~A)  
\begin {eqnarray*}
da(r_{A})=\int d\mu (\int_{A}^{B_{1}}d\tau f_{rad}(AB_{1})exp[-\tau]+\\
\int_{B_{1}}^{B_{2}}d\tau f_{rad}(B_{2}B_{1})exp[-\tau]+ \nonumber\\
\end {eqnarray*}

\begin{equation} 
+...+\int_{B_{k}}^{C}d\tau f_{rad}(B_{k}C)exp[-\tau]
\end {equation}

In order to investigate the upward mixing of $^{56}$Ni , we
modified the s15s7b2f model results so as to place radioactive
nickel beyond 750~km/s. The $^{56}$Ni density profile we invoked is
represented by a constant density profile up to velocity v$_{c}$
followed by a power law profile of index $k$ (Figure 8). The total $^{56}$Ni
mass was always held at the s15s7b2f value {\it viz.} 0.0643
M$_{\odot}$. In each shell the nickel was homogeneously mixed with the
other elements. Therefore a constant value for $f_{rad}$ for
each shell could be determined and the dimensionless deposition
function could be estimated through equation (13). 

In order to calculate the rate of the energy deposited per sec per
cm$^{3}$ from the dimensionless deposition function estimated above we
must take into account the local density of the absorbing material and
the rate of the $\gamma$ ray output,$s_{\gamma}$  (Ambwani \&
Sutherland 1988). We therefore have
\begin {equation}
\epsilon(r)=da(r)  \rho(r)  s_{\gamma}
\end{equation}
where 
\begin {equation}
s_{\gamma}=3.9\times
10^{10}e^{(-\frac{t}{\tau_{1}})} +  6.78\times10^{9}
[e^{-\frac{t}{\tau_{2}}}-e^{-\frac{t}{\tau_{1}}}] 
  ergs g_{rad}^{-1}s^{-1} 
\end{equation}
where $\tau_{1}$ and $\tau_{2}$ are the decay times of $^{56}$Ni and $^{56}$Co.

\newpage
\begin{table}
\footnotesize
\caption{Log of optical spectroscopy of SN~1995V}
\begin{center}
\begin{tabular}{cccccc}
{\bf Date UT 1995}  &{\bf Epoch} $\mathbf({d)}$&
{\bf Telescope/Instrument}
& $\mathbf{\lambda\lambda (\AA)}$ &$\mathbf{\Delta\lambda (\AA)}^{\it
\alpha}$ &{\bf Slit} (arcsec) \\
\hline
Aug 17.2 & 23 & INT/IDS & 3994-9256 & 6.5 & 1.5 \\
Aug 28.2 & 34 & INT/IDS & 4165-9283 & 6.5 & 1.4 \\
Oct 4.1  & 72 & INT/IDS & 3700-9522 & 6.2 & 1.5 \\
Oct 16.1 & 84 &INT/IDS  & 3950-9250 & 6.3 & 1.6 \\
\hline
\end{tabular}
\end{center}
\scriptsize
\baselineskip 0.2cm
\hspace*{1.5cm}
$^{\it \alpha }$Spectral resolution. 
\end{table}

\begin{table}
\caption{ Uncorrected B and V magnitudes derived from the spectra of SN~1995V }
\begin{center}
\begin{tabular}{ccc}
{\bf Epoch} (d) &  {\bf B }& {\bf V} \\
\hline
23   & 16.45  & 15.86  \\  
34   & 16.70  & 15.80  \\
72   & 17.20  & 15.84  \\
84   & 17.47  & 16.24   \\
\hline
\end{tabular}
\end{center}
\end{table} 

\begin{table}
\footnotesize
\caption{Log of infrared spectroscopy of SN~1995V}
\begin{center}
\begin{tabular}{cccccc}
{\bf Date UT 1995}  &{\bf Epoch} $\mathbf({d)}$&
{\bf Telescope/Instrument}
& $\mathbf{\lambda\lambda (\AA)}$ &$\mathbf{\Delta\lambda (\AA)}^{\it
\alpha}$ &{\bf Slit} (arcsec) \\ 
\hline
Aug 15.8 & 22 & AAT/IRIS  & 8416-15149 & 20.7 & 1.4 \\
Sep 4.6 & 42 & UKIRT/CGS4 & 18766-25427 & 25 & 1.2 \\
Sep 6.6 & 44 & UKIRT/CGS4 & 9988-13365 \&  &  13.5 \&  & 
2.4 \\
 & & & 14400-21122 & 26.5 & \\
Oct 1.6 & 69  & UKIRT/CGS4  & 9820-13310 \&  &  13.5 \&  
& 2.4 \\
 & & & 18800-25373 &  27.5 7 \\
Oct 17  & 85  & AAT/IRIS   &  8441-15150 & 22 & 1.4 \\
\hline
\end{tabular}
\end{center}
\scriptsize
\baselineskip 0.22cm
\hspace*{1.5cm}
$^{\it \alpha }$Spectral resolution. 

\end{table}

\begin{table}
\normalsize
\caption{Model parameters giving the best reproduction of
the observed infrared and optical spectra (optically thin lines only).}
\begin{center}
\begin{tabular}{ccc}

  {\bf Epoch} (d)& {$\bf v(ph)$} (km/s)& {$\bf n$} \\ 
\hline
   22-23 & 6200  & -5.1  \\
   44-34 & 3700  & -4.5  \\
   69-72 & 2600  & -3.0  \\
   85-84 & 2200  & -3.0  \\
\hline
\end{tabular}
\end{center}
\end{table}

\newpage

\begin{table}
\footnotesize
\caption{Line identifications and derived velocities, v(ph),
~(km/s) and lower level populations, N$_{low}(ph)$, ~(cm$^{-3}$) for the
optical spectra}
\begin{center}
\begin{tabular}{ccccccccc}
{} &\multicolumn{2}{c}{\bf 23 days} &\multicolumn{2}{c}{\bf 34 days} 
&\multicolumn{2}{c}{\bf 72 days} &\multicolumn{2}{c}{\bf 84 days}\\ \hline
{\bf Line}&{$\bf v(ph)$}&{$\bf N_{low}(ph)$}&{$\bf v(ph)$}&{$\bf N_{low}(ph)$}&
{$\bf v(ph)$}&{$\bf N_{low}(ph)$}
&{$\bf v(ph)$}&{$\bf N_{low}(ph)$} \\ \hline 
\small
Sr~II~4077 &{-}  & {-}   &   {-}  & {-}  & 2850 &  1.1  & 2450  & 0.7  \\
H$\delta$  &{-}  &  {-}  &   {-}  & {-}  & 3100 &  0.6  & 2800  & 1.1 \\
Ca~I~4142  &{-}  &  {-}  &   {-}  & {-}  & 2600 &  0.7 & 2200  & 0.4 \\
Ba~II~4166 &{-}  &  {-}  &   {-}  & {-}  & 2600 &  3.5  & 2200  & 3.0  \\ 
Fe~II~4205 &{-}  &  {-}  &   {-}  &  {-}  & 2600 &  8.5   & 2200  & 7.5   \\
Sr~II~4215 &{-}  &  {-}  &   {-}  &  {-}  & 2500 &  0.2 & 2200  & 0.22 \\
Ca~II~4226 &{-}  &  {-}  &  3700  &  0.34  & 2500 &  0.23 & 2200  & 0.22 \\
Ca~II~4289 &{-}  &  {-}  &  4100  &  0.95 & 2600 & 0.6  & 2200  & 0.5 \\
H$\gamma$  &7100 &  17.5 &  5350  &  16.5  & 3100 &  13.5  & 2800  & 8.5  \\ 
Ca~I~4355  &{-}  &  {-}  &  4000  &  15.5  & 2600 &  1.6  & 2200  & 1.3 \\ 
Sc~II~4400 &{-}  & {-}   &  {-}   &  {-}  & 2600 & 7.5 & 2200  & 10.0  \\ 
Ca~I~4425  &{-}  & {-}   &  {-}   &  {-}  & 2600 & 10.1 & 2200  & 9.0 \\
Ba~II~4524 &{-}  & {-}   &  3850  &  1.8  & 2600 &  1.25  & 2200  & 0.9 \\
Ba~II~4554 &6400 & 1.3   &  3850  &  2.0  & 2600 &  1.12  & 2200  & 0.9 \\
Ca~I~4578  &{-}  &  {-}  &   {-}  &  {-}  & 2600 &   5   & 2200  & 3  \\
Ca~I~4581  &{-}  &  {-}  &   {-}  &  {-}  & 2600 &   1.33   & 2200  & 1.9  \\
Ca~I~4585  &{-}  &  {-}  &   {-}  &  {-}  & 2600 &  0.28 & 2200  & 0.3 \\
C~II~4627  &{-}  &  {-}  &   {-}  &  {-}  & 2500 &  12.3  & 2100  & 7 \\
Sc~II~4670 &{-}  &  {-}  &   {-}  &  {-}  & 2700 &   3 & 2200  & 8 \\
H$\beta$   &7000 &  8    &  5100  &  6.6 & 3500 & 3.5  & 2900  & 2.05 \\
Ba~II~4934 &{-}  &  {-}  &  4800  &  1.65  & 2850 &  2.9  & 2450  & 2.9 \\
Fe~II~5018 &6200 &  49   &  4400  & 115   & 2700 &  70  & 2400  & 65  \\
Fe~II~5169 &6000 &  38.5 &  4350  &  52   & 2400 &  45   & 2050  & 47 \\
Fe~II~5322 &{-}  &  {-}  &  4700  &  5.0  & 2600 &   6 & 2400  & 3.4  \\
Fe~II~5429 &{-}  &  {-}  &   {-}  &  {-}  & 2600 &  0.3 & 2400  & 0.1 \\
Sc~II~5526 &{-}  &  {-}  &   {-}  &  {-}  & 2600 &  1.13 & 2200  & 0.40 \\
Sc~II~5657 &{-}  & {-}   &   {-}  &  {-}  & 2650 &  2.8  & 2100  & 1.65 \\
Ba~II~5853 &{-}  &  {-}  &   {-}  &  {-}  & 2750 &  7.9  & 2500  & 5.5 \\
Na~I~5889  &{-}  &  {-}  &  4200  &  0.15  & 2950 &  0.33 & 2500  & 0.3 \\
Na~I~5895  &{-}  & {-}   &   {-}  &  {-}  & 2950 &  0.33 & 2500  & 0.3 \\
Ba~II~6142 &{-}  & {-}   &   {-}  &  {-}  & 2600 &  0.27 & 2250  & 0.32 \\
Sc~II~6245 &{-}  & {-}   &   {-}  &  {-}  & 2650 &  0.9   & 2200  & 0.65 \\ 
Sc~II~6378 &{-}  & {-}   &   {-}  &  {-}  & 2600 &  0.07 & 2200  & 0.1 \\
Ba~II~6497 &{-}  & {-}   &   {-}  &  {-}  & 2600 &  0.35 & 2150  & 0.33\\
H$\alpha$  &8500 & 0.7  &  6650  &  0.8  & 4300 &  0.95 & 3850  & 0.85 \\
Ca~II~8498 &  6550 &  49  &  5300  &  55 & 3700 &  17  & 3300  & 21\\
Ca~II~8542 &  6550 &  1.85   &  5300  &  1.65 & 3700 &  0.6 & 3300  & 0.47 \\
Ca~II~8662 &  6400 &  12  &  5300  &  11.5  & 3700 &  9 & 3200 & 9 \\
C~I~9088   &{-}  &  {-}  &  3800  & 0.3  & 2450 &  0.18 &  {-}  &  {-}\\ 
Sc~II~9236 &{-}  &  {-} &   {-}  &  {-}  & 2600 &  0.23 &  {-} & {-}\\  
C~I~9405    &{-} &  {-}  &   {-}  &  {-} & 2600 &  0.23 &  {-}  &  {-}\\
\hline
\end{tabular}
\end{center}
\end{table}

\begin{table}
\caption{ Line identifications and derived velocities, v(ph),
~(km/s) and lower level populations, N$_{low}(ph)$, ~(cm$^{-3}$) for the
infrared spectra} 
\begin{center}
\begin{tabular}{ccccccccc}
{} &\multicolumn{2}{c}{\bf 22 days} &\multicolumn{2}{c}{\bf 44 days} 
&\multicolumn{2}{c}{\bf 69 days} &\multicolumn{2}{c}{\bf 85 days}\\ \hline
{\bf Line}&{$\bf v(ph)$}&{$\bf N_{low}(ph)$}&{$\bf v(ph)$}&{$\bf
N_{low}(ph)$}&{$\bf v(ph)$}&{$\bf N_{low}(ph)$}
&{$\bf v(ph)$}&{$\bf N_{low}(ph)$} \\ \hline 
P$\delta$ & 6200 & 2.1 & 4300  & 0.3  & 3000  &  0.1    & 2350   & 0.3  \\
Sr II 10327  &  {-} &{-}& 3300  &  0.23  & 2600  &  0.6 & 2350  &1.15 \\
Fe II 10547  &  {-} &{-}& 3300  &  1.65  & 2450  &  0.5  & 2200 & 0.45 \\
C~I~10695    &  6500 &0.5& 3700  & 0.17  & 2350  &  0.08 & 2200  & 0.1 \\ 
He~I~10830   &  {-}  &{-}& 6000  &  0.055 & 4250 & 0.21  &  3750 & 0.23 \\
P$\gamma$    &  6100 & 1.25& 4200 &  0.16 & 3200 &  0.27 & 2600 & 0.7 \\
P$\beta$     &  6100 & 0.2 & 4200  & 0.15  & 3700& 0.05  & 2700 &0.03  \\ 
P$\alpha$    & {-} & {-}& 4700 & 0.04 & 3700 &0.08  & {-} &{-}  \\ 
Br$\gamma$ &{-} & {-}& 4100 & 0.08 & 3000 & 0.04& {-} & {-} \\
\hline
\end{tabular}
\end{center}
\end{table}

\begin{table}
\caption{Rates for the depopulation processes of the 2s$^{3}$S level.}
\begin{center}
\begin{tabular}{cc}
  {\bf Depopulation process} & {\bf Rate} s$^{-1}$ \\ 
\hline
   Radiative decay &  1.24$\times$ 10$^{-4}$ \\
   Collisions to singlet states & 1.8$\times$ 10$^{-2}$  \\
2s$^{3}$S$\rightarrow$2p$^{3}$P$\rightarrow$1s$^{1}$S & 4$\times$ 10$^{-3}$ \\
   Penning ionisation & 92     \\
\hline
\end{tabular}
\end{center}

\end{table}

\begin{table}
\normalsize
\caption{ Results of the modelling and $^{56}$Ni mass estimates}
\begin{center}
\begin{tabular}{ccccc}
 $\bf \chi_{He}$ & ${\bf k}$ &{\bf v$_{c}$} (km/s)
& {\bf $^{56}$Ni mass in the power }&{\bf $^{56}$Ni mass above the helium} \\
& & &{\bf law zone} ($\times$ 10$^{-2}$ M$_{\odot}$)&
{\bf photosphere}($\times$ 10$^{-7}$ M$_{\odot}$) \\
\hline
   0   &  8$\pm$1 & 1150$\pm$ 300 &2.43 $\pm$ 0.25 & 750 $\pm$ 26\\
   0.1 &  9.5$\pm$1.5 & 880$\pm$350 &2.05 $\pm$ 0.24 & 19  $\pm$ 1.5\\
   0.2 &  9$\pm$1 & 650$\pm$200 &2.1 $\pm$ 0.18 & 6.3 $\pm$ 0.43 \\
\hline
\end{tabular}
\end{center}
\end{table}
\clearpage
\newpage
\bf\Large{Figure Captions}\\
\\ 
\normalsize
{\bf Figure 1:} Optical spectra of SN~1995V taken with the Isaac Newton
Telescope, La Palma (see Table~1 for details). The fluxes have been
scaled as explained in the text. For clarity the spectra have been
displaced vertically by the amounts indicated.  The days shown are with
respect to the estimated date of shock breakout, 25 July 1995.

{\bf Figure 2:} Comparison of the B~light curve of the type~IIp SN~1983~K (Phillips
{\it et al.} 1990) with
the uncorrected B~magnitudes for SN~1995V derived from the spectra.
To aid the comparison, the SN~1995V points have been vertically
displaced by -3.5 magnitudes. The upper limit of Evans (1995) on 25
July 1995 is also indicated. This, together with the four
B~magnitudes, enables us to fix the time of shock break-out, t$_{0}$,
to a precision of $\pm$2~days.

{\bf Figure 3:} Infrared spectra of SN~1995V taken with the Anglo-Australian
Telescope (days~22 and 85) and the United Kingdom Infrared Telescope
(days 42, 44 and 69) (see Table~3 for details).  The fluxes have been
scaled as explained in the text.  For clarity the spectra have been
displaced vertically by the amounts indicated. The days shown are with
respect to the estimated date of shock breakout, 25 July 1995.

{\bf Figure 4:} Observed  optical spectra of SN~1995V compared with the spectral
model.  Details of the model parameters, line velocities and
populations of the transition lower levels are presented
in Tables 4 and 5. Absorption minima of identified lines are indicated.

{\bf Figure 5:} Observed infrared spectra of SN~1995V compared with the spectral
model.  Details of the model parameters, line velocities and
populations of the transition lower levels are presented
in Tables 4 and 6.  Absorption minima of identified lines are indicated.  

{\bf Figure 6:} Comparison of two model fits for 69 and 85 days.
The dashed line represents the model fit when He~I~10830~\AA\ is
{\it not} included.  The model fit which includes He~I~10830~\AA\
(solid strong line) is clearly superior.

{\bf Figure 7:} Synthetic spectra produced using the $^{56}$Ni density profile
of the explosion model s15s7b2f compared with observations.
The quantity $\chi_{He}$ indicates the fraction of the helium mass that
was in the form of clumps for each model.

{\bf Figure 8:} The $^{56}$Ni density profile used to represent the
dredge-up of the radioactive material

{\bf Figure 9:} $^{56}$Ni density profiles that produce the best fit synthetic
spectra. (The helium photosphere lies at $\sim$ 4250 km/s for day 69)

{\bf Figure 10:} Synthetic spectra produced using the  $^{56}$Ni profile of
figure 8 with parameters $k=9.0$, v$_{c}=800km/s$ and assuming that
10\% of the helium mass was in the form of clumps.
\newpage

\begin{figure}
\centerline{\psfig{figure=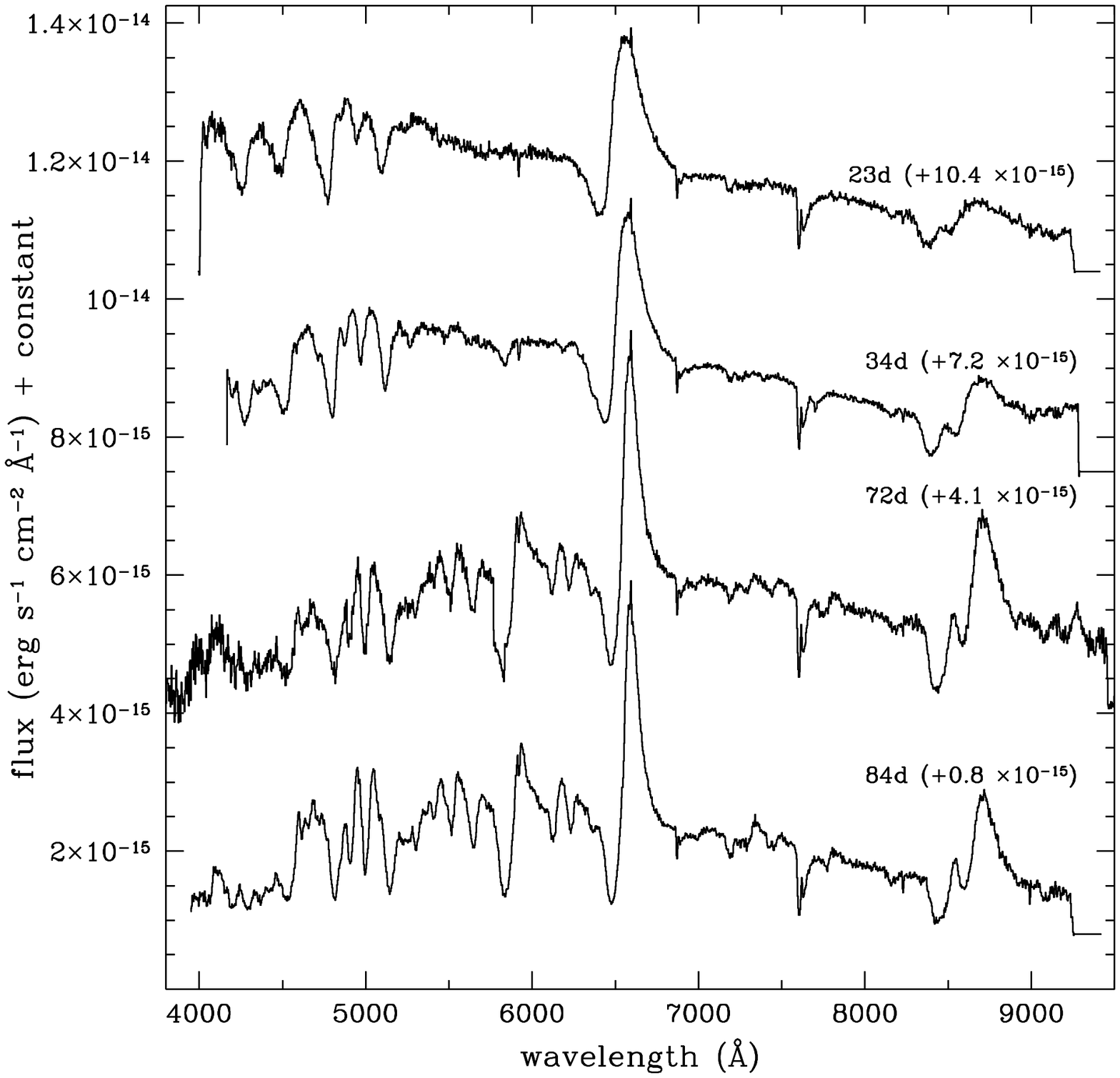,width=6.1in,height=6.0in,angle=0}}
\caption{}
\end{figure}

\begin{figure}
\centerline{\psfig{figure=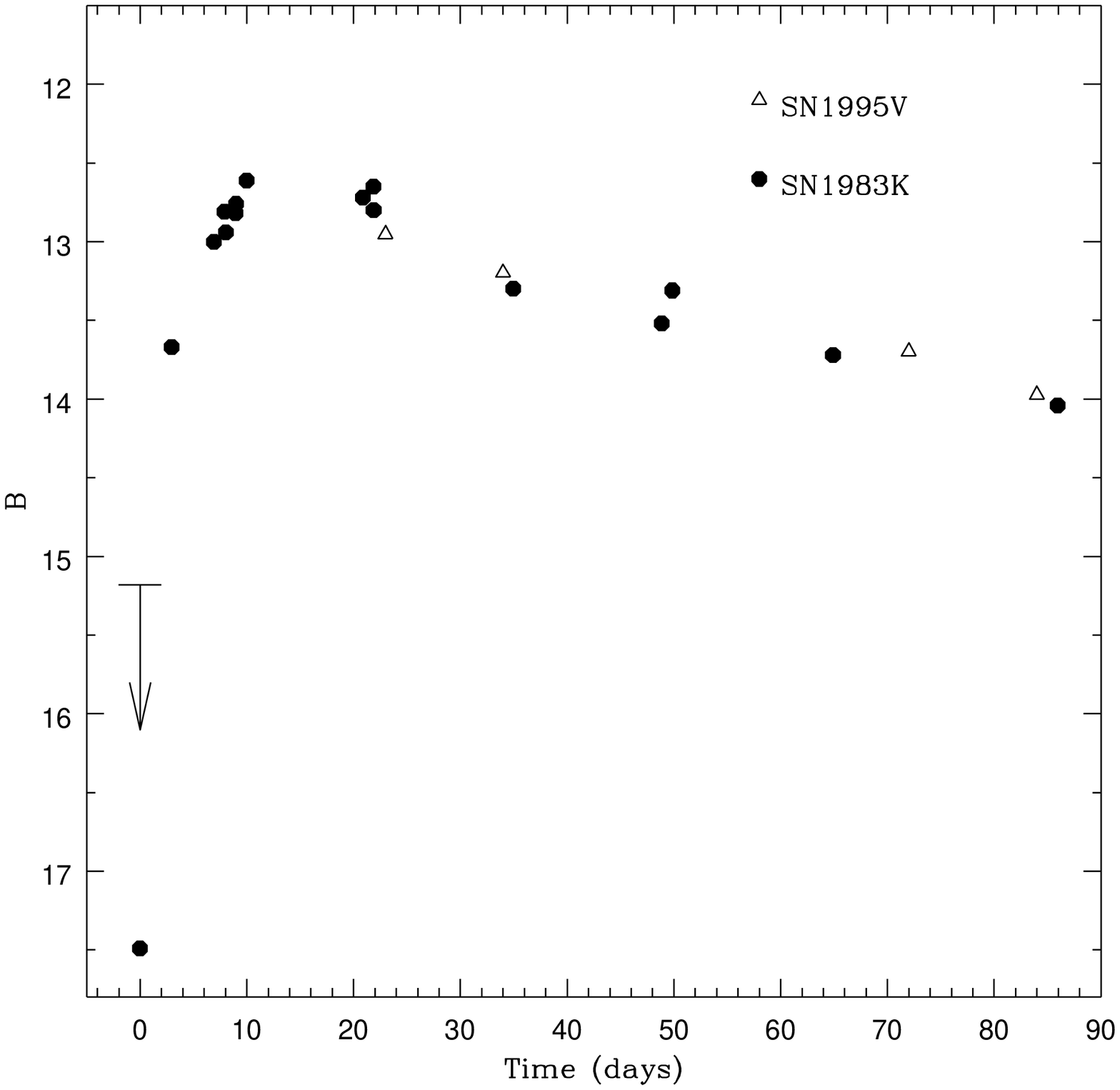,width=6.1in,height=6.0in,angle=0}}
\caption{}
\end{figure}

\begin{figure}
\centerline{\psfig{figure=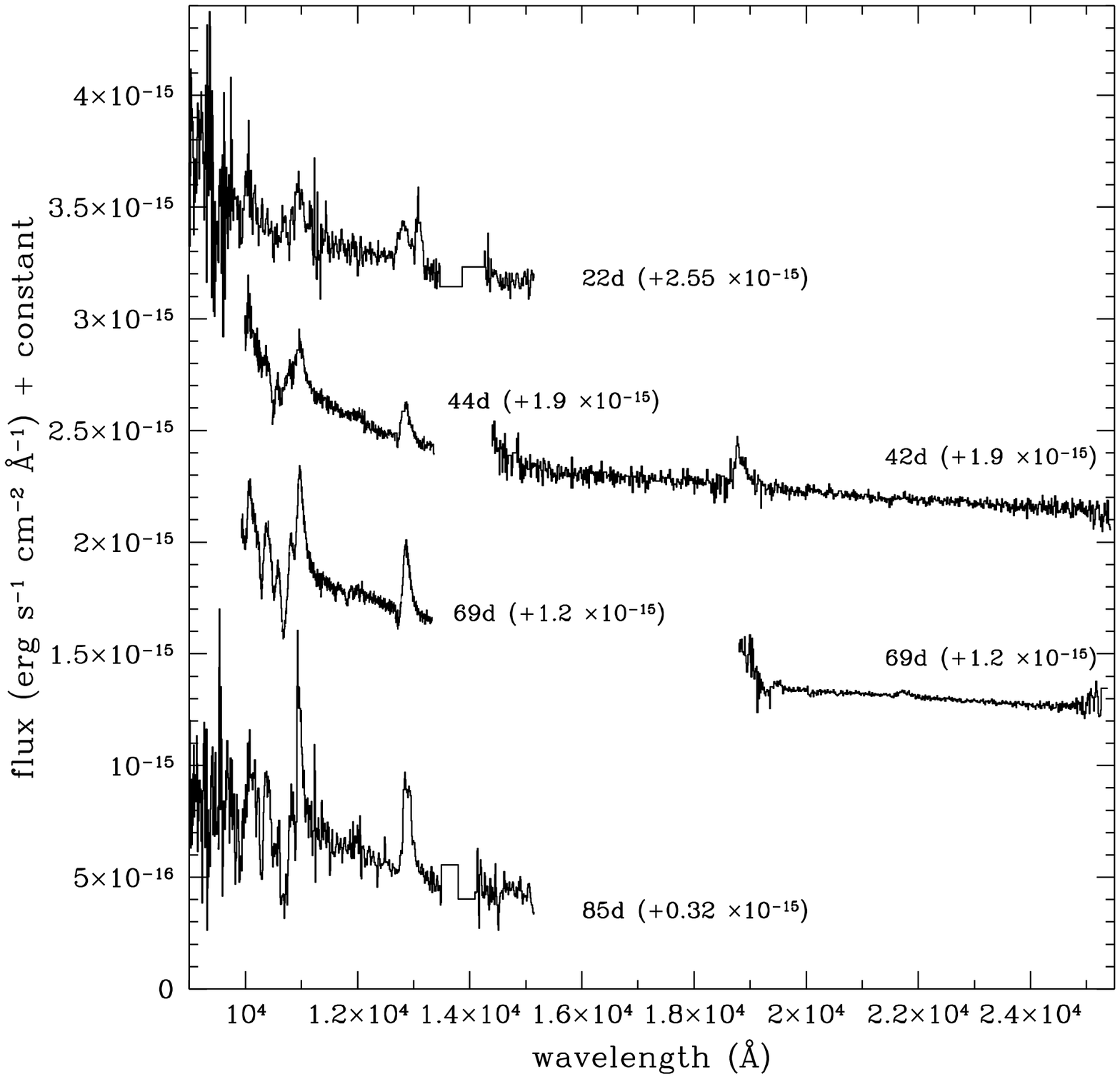,width=6.1in,height=6.0in,angle=0}}
\caption{}
\end{figure}

\begin{figure}
\centerline{\psfig{figure=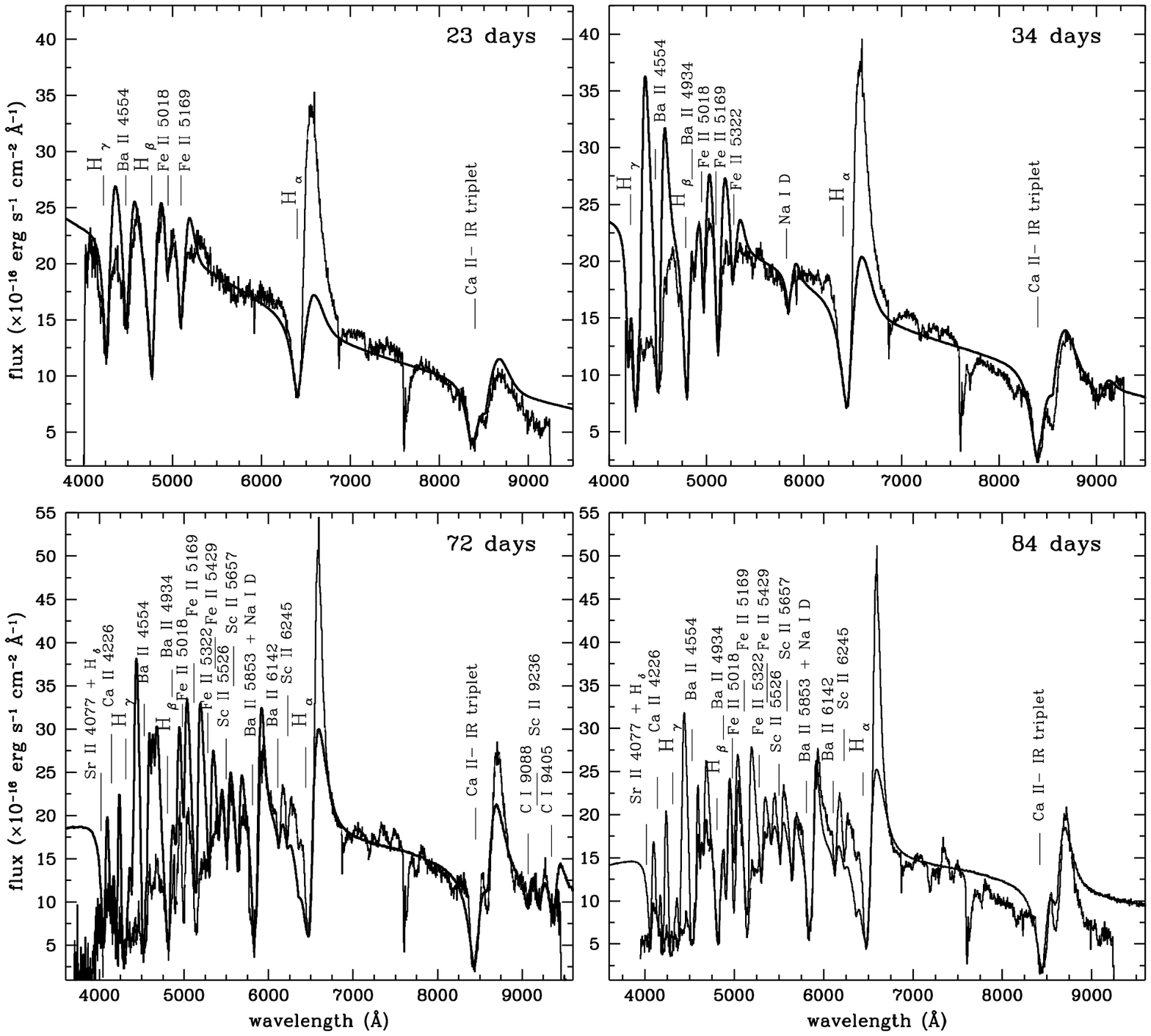,width=7.2in,height=8.5in,angle=0}}
\caption{}
\end{figure}

\begin{figure}
\centerline{\psfig{figure=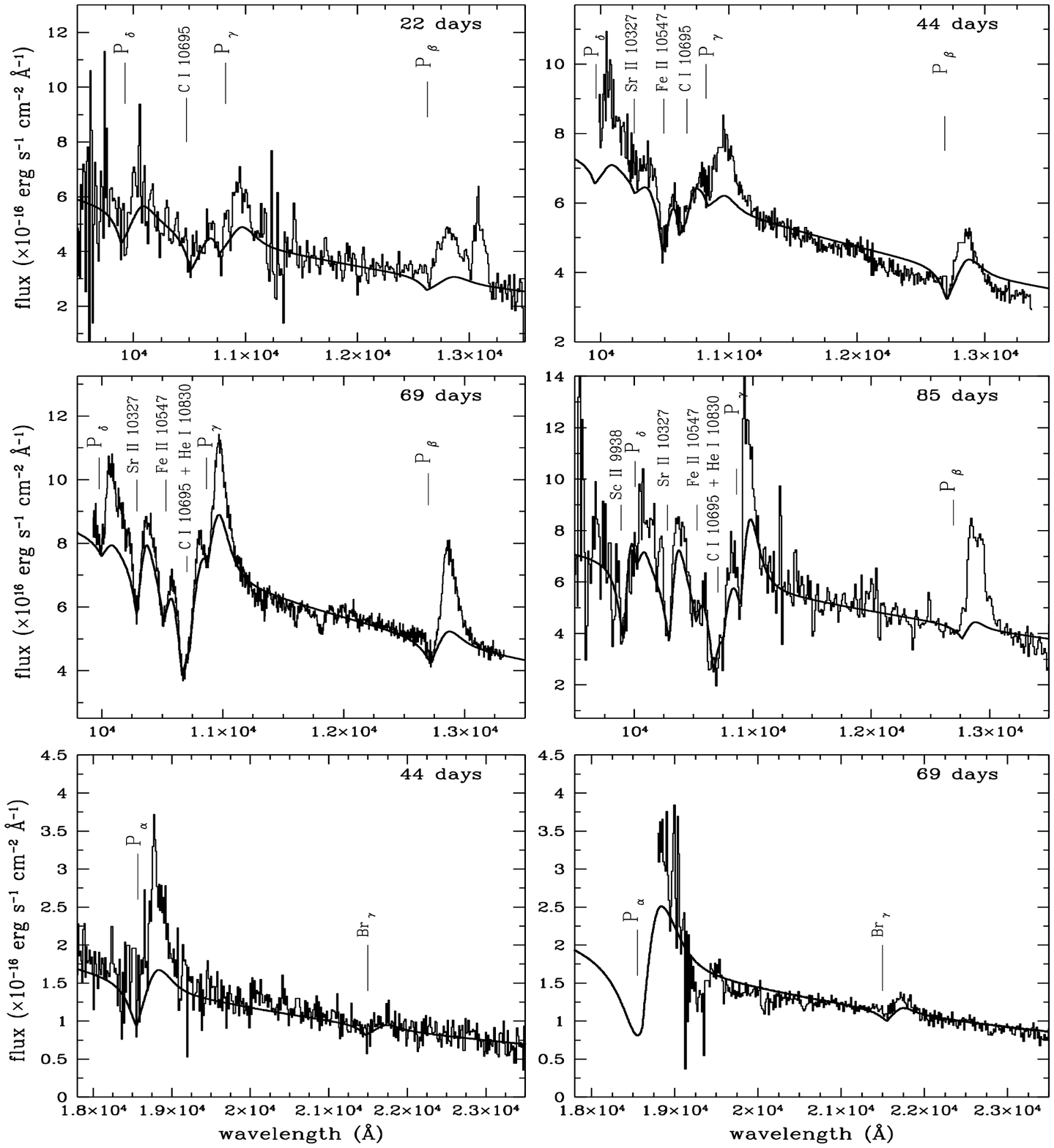,width=7.2in,height=8.5in,angle=0}}
\caption{}
\end{figure}

\begin{figure}
\centerline{\psfig{figure=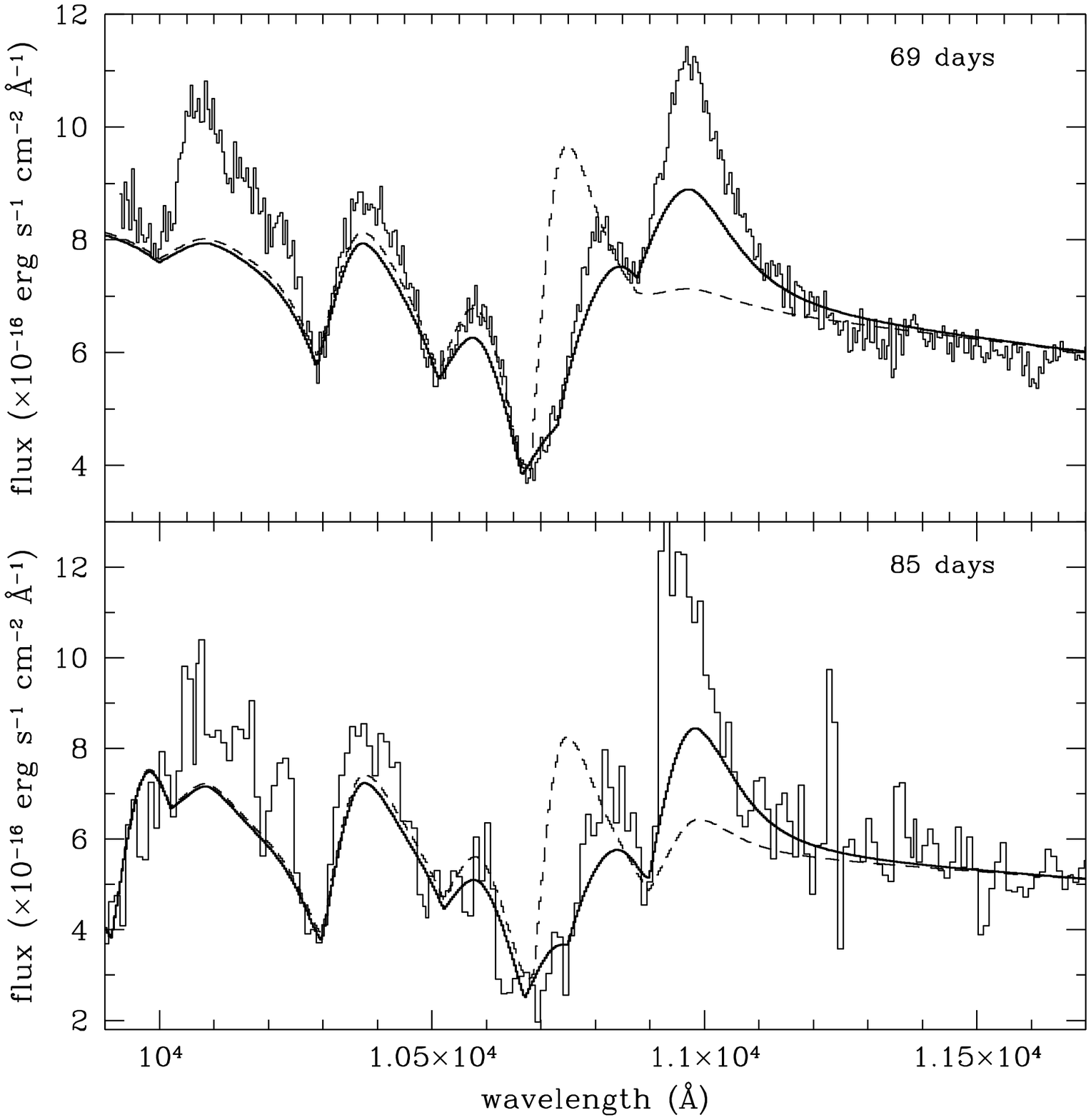,width=5.8in,height=5.8in,angle=0}}
\caption{}
\end{figure}

\begin{figure}
\centerline{\psfig{figure=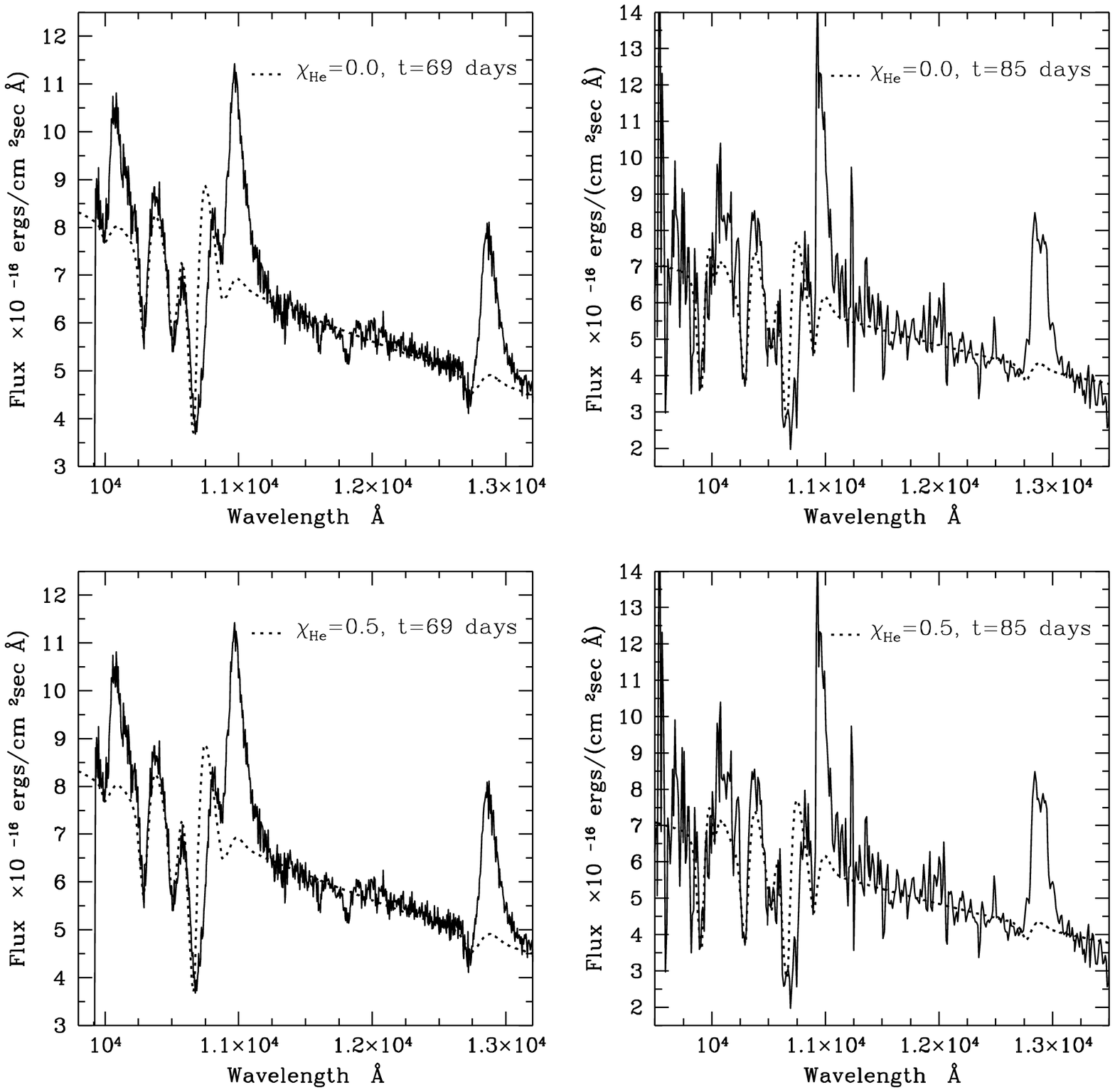,width=6.5in,height=6in,angle=0}}
\caption{}
\end{figure}

\begin{figure}
\centerline{\psfig{figure=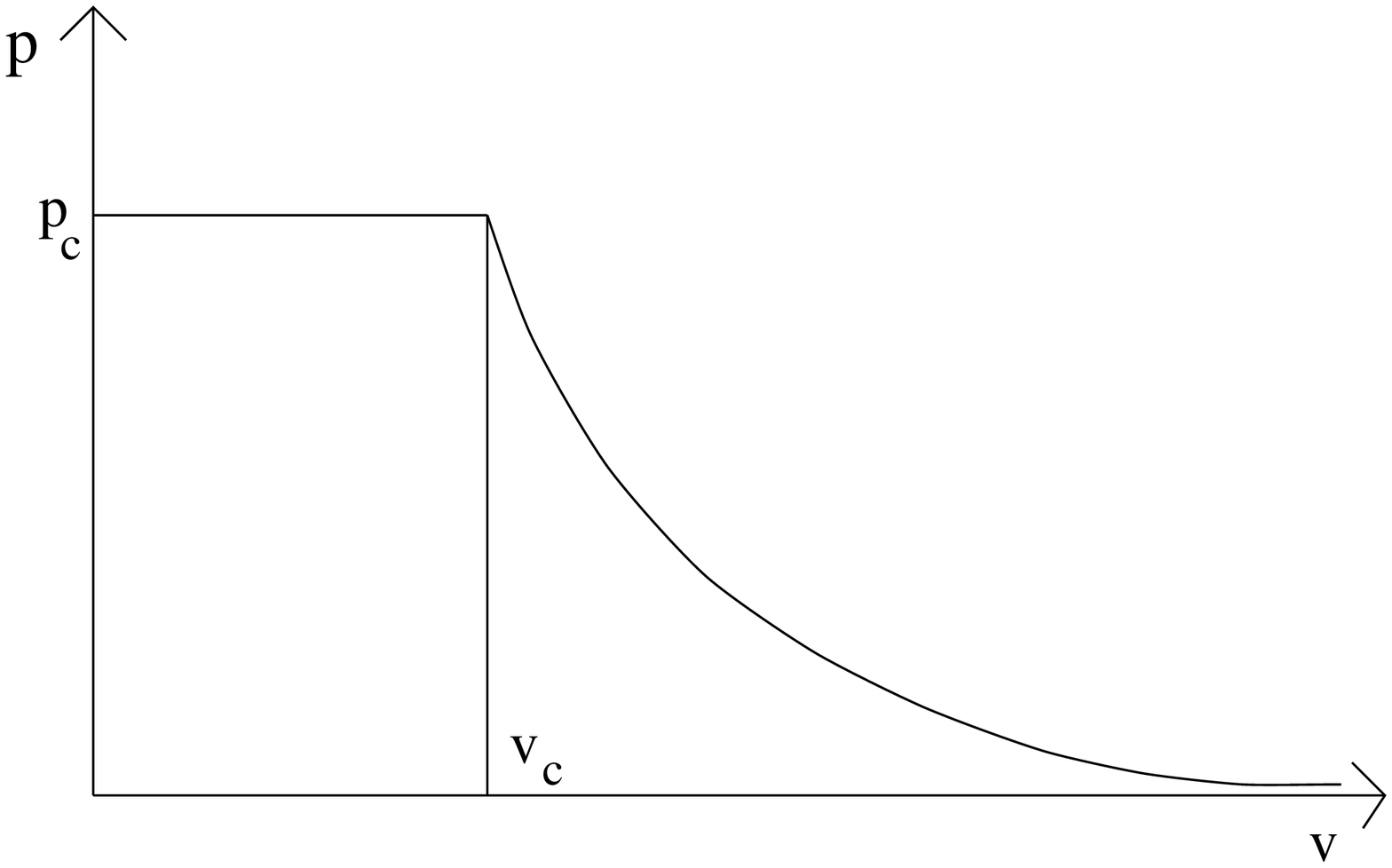,width=3in,height=2in,angle=270}}
\caption{}
\end{figure}

\begin{figure}
\centerline{\psfig{figure=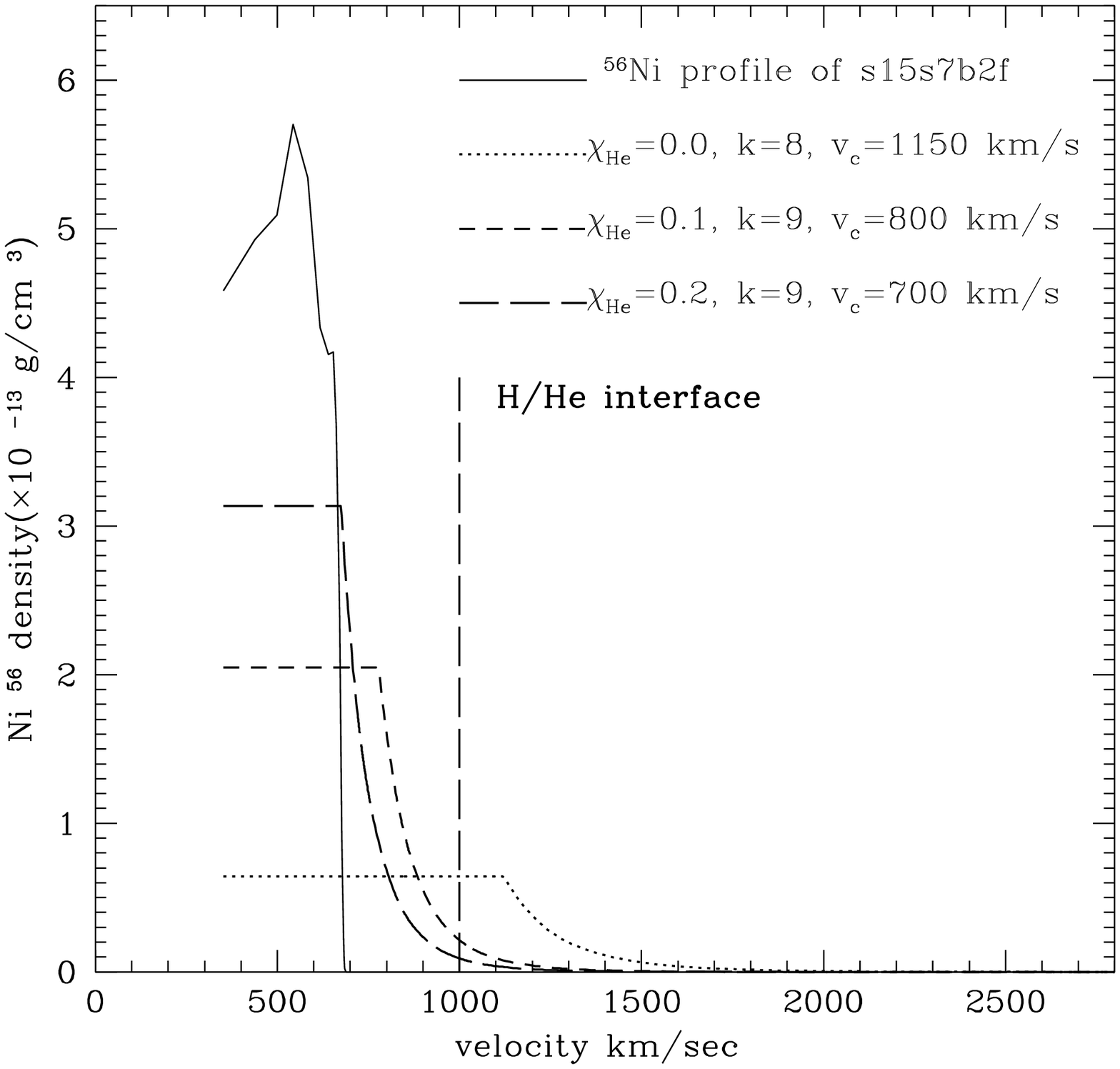,width=5.8in,height=5.8in,angle=0}}
\caption{}
\end{figure}

\begin{figure}
\centerline{\psfig{figure=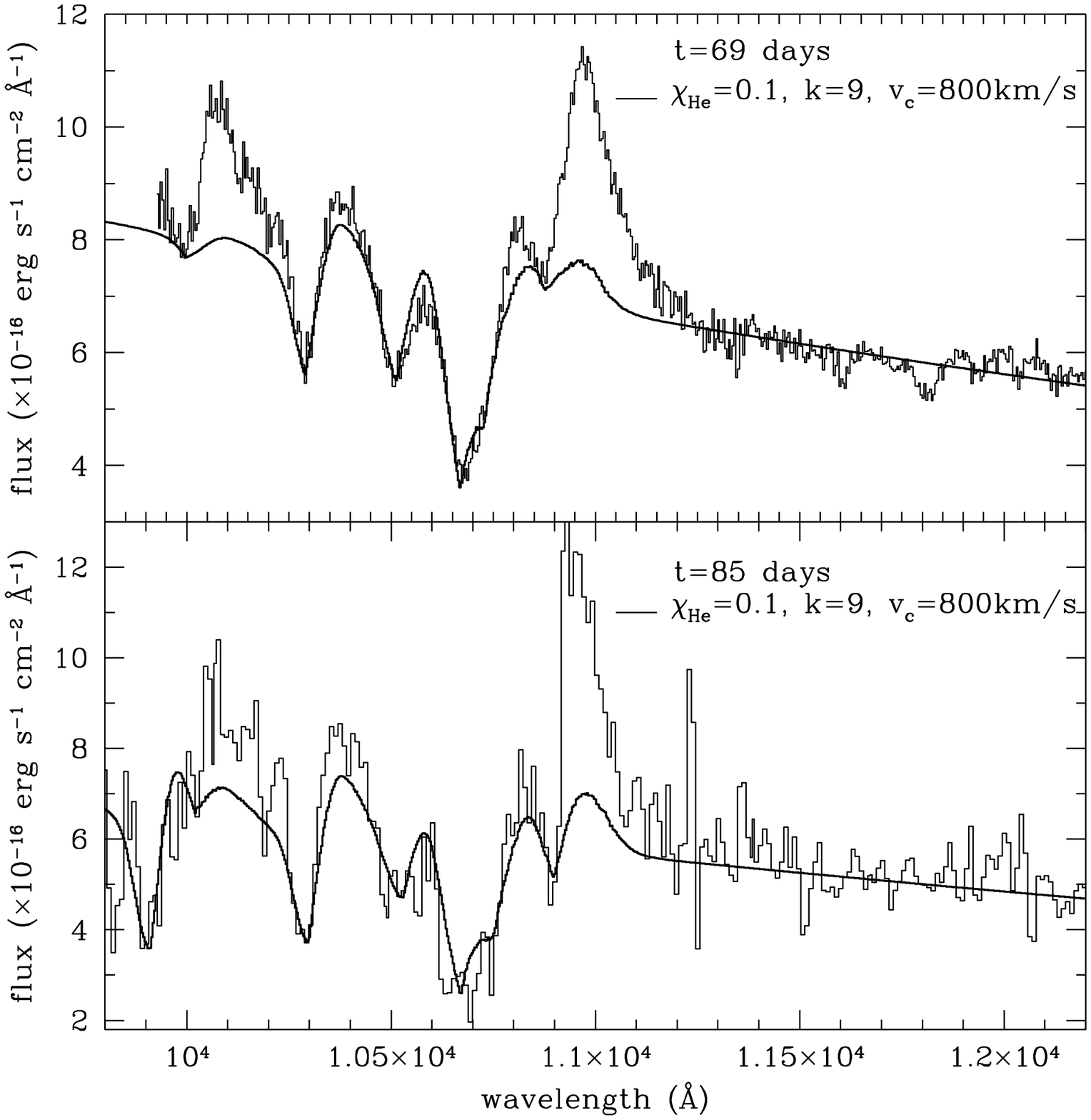,width=5.8in,height=5.8in,angle=0}}
\caption{}
\end{figure} 

\end{document}